\documentclass[aps,twocolumn,pra,superscript,floatfix,superscriptaddress,showpacs,footinbib]{revtex4-1}

\usepackage[english]{babel}
\usepackage{amsmath}
\usepackage{multirow}
\usepackage{bm}
\usepackage{graphicx}
\usepackage[colorlinks=true,
            linkcolor=blue,
            urlcolor=blue,
            citecolor=blue,
            filecolor=blue,
            bookmarksopen=true,
            pdfauthor={J. P. Ramos-Andrade},
            pdftitle={class_measurement},
            pdfsubject={class_measurement_Manuscript},
            pdfpagemode=UseOutlines]{hyperref}

\graphicspath{{Pictures/}}

\begin{document}

\title{Josephson and persistent currents in a quantum ring between topological superconductors}
\author{F. G. Medina}
\author{J. P. Ramos-Andrade}
\author{L. Rosales}
\author{P. A. Orellana}
\affiliation{Departamento de F\'isica, Universidad T\'ecnica Federico Santa Mar\'ia, Casilla 110 V, Valpara\'iso, Chile}

\begin{abstract}
In this work, we investigate the spectra in an Aharonov-Bohm quantum-ring interferometer forming a Josephson junction between two topological superconductor (TSC) nanowires. The TSCs host Majorana bound states at their edges, and both the magnetic flux and the superconducting phase difference between the TSCs are used as control parameters. We use a tight-binding approach to model the quantum ring coupled to both TSCs, described by the Kitaev effective Hamiltonian. We solve the problem by means of exact numerical diagonalization of the Bogoliubov-de Gennes (BdG) Hamiltonian and obtain the spectra for two sizes of the quantum ring as a function of the magnetic flux and the phase difference between the TSCs. Depending on the size of the quantum ring and the coupling, the spectra display several patterns. Those are denoted as line, point and undulated nodes, together with flat bands, which are topologically protected. The first three patterns can be possibly detected by means of persistent and Josephson currents. Hence, our results could be useful to understand the spectra and their relation with the behavior of the current signals.
\end{abstract}

\maketitle

\section{Introduction}

Topological properties in matter have been intensively investigated in the past decade in different contexts, starting with gapped topological phases in insulators, superconductors \cite{hasan2010colloquium,qi2011topological,sato2017topological} and in gapless topological phases, such as Weyl semimetals \cite{lu2015experimental,xu2015discovery,soluyanov2015type}. The classification of these phases is summarized in the tenfold way, with time reversal ($\mathcal{T}$), particle-hole ($\mathcal{PH}$) and Chiral ($\mathcal{C}$) symmetries being the cornerstone of topological matter \cite{chiu2016classification,kotetes2013classification,schnyder2008classification}. One widely studied system is the Kitaev chain that belongs to class D, since it is a system with broken $\mathcal{T}$ and $\mathcal{C}$ symmetries, leaving the $\mathcal{PH}$ symmetry as the one that protects the edge modes in the Kitaev chain in its topological phase. These modes, also known as Majorana bound states (MBSs), \cite{kitaev2001unpaired, alecce2017extended, miao2017exact} receive important attention in their detection, manipulation and control with the objective of being implemented in quantum computing \cite{matos2017tunable, alicea2011non,sarma2015majorana,van2012coulomb}. Originally, the idea of the Kitaev chain had a complication, which relies on the fact that to possess the topological phase, the fermionic states must be spinless fermions. This complication has been overcome since spinless fermions can now be understood experimentally as fully polarized fermions, and under the effect of superconductivity and high Rashba spin-orbit interaction, spin-triplet superconductors are produced \cite{mourik2012signatures, deng2012anomalous}. 

Intrinsic spin-triplet superconductors (SCs) have been found in different materials, such as Li$_2$Pt$_3$B due to its broken inversion symmetry \cite{nishiyama2007spin} and in quasi one-dimensional chromium pnictide K$_2$Cr$_3$As$_3$ \cite{bao2015superconductivity}. But they have also been produced by artificially implementing spin-singlet SCs: in interfaces of SC-ferromagnet \cite{eschrig2015spin}, or by employing the most common setup for observing MBSs, which consists of taking Al-InAs or InSb nanowires and inducing superconductivity by the proximity effect.  \cite{das2012zero,dominguez2017zero,lutchyn2018majorana}.        

In this context, Josephson junction devices with spin-singlet pairing SCs overcome topological characteristics that go beyond the tenfold classification, displaying Weyl points in the parameter space \cite{meyer2017nontrivial, xie2017topological,riwar2016multi,xie2018weyl}. The classification of these hybrid systems could be clarified following the work of Zhang et al. \cite{zhang2014anomalous}. In recent works, multiterminal Josephson junctions with spin-triplet paring SCs hosting MBSs at the edges provide a twist in these reputed Weyl points. Since Weyl points in crystals require the breaking of $\mathcal{T}$ or inversion symmetry, in four-terminal junctions, the Weyl and nodal points may also occur under $\mathcal{T}$ symmetry \cite{sakurai2020nodal,houzet2019majorana,kotetes2019synthetic}. The importance of the Josephson junctions lies in the engineering of Andreev bound states (ABSs) and their control by building topological quantum-computing architectures based on MBSs \cite{nayak2008non,PhysRevB.84.094505,huang2015manipulating, wang2018landau,trif2018dynamic}, where superconducting phase difference, bias voltage, and magnetic flux play the role of control parameters \cite{malciu2018braiding,hyart2013flux}. To observe MBSs in 1D topological superconductors (TSCs), one of the most suggested models is based on quantum dots (QDs) because they are easily tunnelable objects. Certainly, there are proposals based on observing the half-integer conductance at zero bias \cite{vernek2014subtle,liu2011detecting}, or the nonlocality of MBSs in 1D hybrid nanowires \cite{deng2018nonlocality}, which is an experimental observation. Some theoretical proposals are based on the structure of an Aharanov-Bohm quantum-ring, where it is suggested to observe the discontinuities in persistent currents to detect zero energy crossing or nodal behaviors \cite{medina2020influence, nava2017persistent} or in the persistence conductance, which also displays nodal behavior \cite{chiu2018conductance}.  

In this work, we investigate a low-energy model that describes a bi-junction consisting of an Aharonov-Bohm quantum-ring between two 1D TSCs. The analysis of the system is made in 2D parameter space, composed of the phase difference between the TSCs and the magnetic flux. The ABS spectra are obtained from the exact numerical diagonalization of the BdG Hamiltonian, observing zero energy crossing or nodes, flat bands, and the coexistence of both. These three features are the topological characteristics of the spectra in a bi-junction that involves four MBSs. Interestingly, some band structures obtained for semimetals \cite{liu2014stable,neupane2014observation} could be studied using the interplay in magnetic flux and phase differences of TSCs. In addition, we found that the persistent and dc Josephson currents can be used to detect those topological patterns. Both current signals are mixed in our system, and our results shows that they can be controlled by tuning the magnetic flux and/or the phase difference.

This work is organized as follows. In Sec. \ref{Sec.2}, we describe the model and the theoretical considerations that we take before diagonalization. In Sec. \ref{Sec.3}, we explain the coupling configurations from the MBSs with the quantum ring and present the spectra. In Sec. \ref{Sec.4}, we present persistent and dc Josephson currents and the interplay between them through the parameter space ($\Phi, \theta/\pi$). Finally, in Sec. \ref{Sec. 5} we summarize the results which are stored in Tables \ref{table1} and \ref{table2}.   
\begin{figure*}[ht]
	\centering
	\includegraphics[width=.6\textwidth]{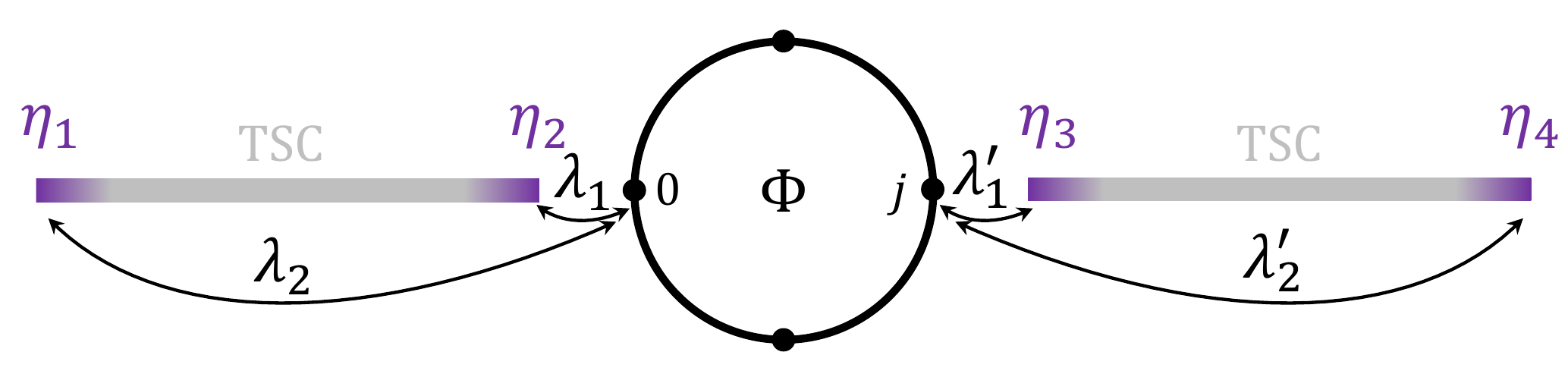}
	\caption{\label{fig:schematic_model} Schematic view of the system under study: Two TSCs (gray) hosting MBSs in their edges (purple) connected with a quantum-ring, crossed by a magnetic flux. The left TSC is coupled with the 0-site, while the other TSC $j$-th site. The couplings setup presented corresponds to Eq.\ (\ref{eq:Coupling Hamiltonian}). Note that when $\lambda_{2}=\lambda_{2}'=0$ and $\lambda_{1}(\lambda_{1}')\rightarrow\lambda(\lambda')$ correspond to Eq.\ (\ref{H5}).}
\end{figure*}

\section{Model and procedure}\label{Sec.2}
\subsection{Model}

The theoretical setup is based on a bi-junction consisting of a tight-binding model for a 1D spinless Aharonov-Bohm quantum ring with two D-class TSCs attached to the side, as it is schematically represented in Fig \ref{fig:schematic_model}. The Hamiltonian for the two TSCs is chosen effectively such that it only captures the edge states. Hence, the Hamiltonian in the Majorana representation is written as:
\begin{equation}
	H_{\text{TSC}} = i\xi_{1}\eta_{1}\eta_{2} + i\xi_{2}\eta_{3}\eta_{4}\,,
	\label{eq:Wire Hamiltonian}
\end{equation}
with $\xi_{1\left(2\right)} \sim e^{-L_{1\left(2\right)}}$ the coupling amplitude proportional to the correlation length between MBSs and $L_{1\left(2\right)}$ the ratio between nanowire length and the superconducting coherence length. Note that we do not consider Majorana oscillations in each TSC. These oscillations appear when Majorana modes, belonging to the same TSC, interact with each other and have a dependence on either the Zeeman field or chemical potential \cite{sarma2012splitting,rainis2013towards,danon2017conductance,ricco2018majorana}. The Hamiltonian for the quantum-ring in the position space is given by:
\begin{equation}
	H_{\text{R}} = \sum^{N - 1}_{l = 0}e^{i\frac{2\pi}{N}\Phi} t c^{\dagger}_{l}c_{l+1} + \text{h.c.}\,,
	\label{eq:Quantum-ring Hamiltonian}
\end{equation}
with $N$ the total number of states in the quantum-ring, $\Phi$ the magnetic flux in units of $\phi_{0} = h/e$ and $t$ the hopping amplitudes between QDs. A useful way to approach this Hamiltonian is to write it in the momentum space by choosing the transformation $c_{l} = \sum_{k}e^{ikl}c_{k}$. Then, $H_{\text{R}}$ is expressed as:
\begin{equation}
	H_{\text{R}} =  \sum_{k}2t\cos\left(k + \frac{2\pi}{N}\Phi\right)c^{\dagger}_{k}c_{k}\,,
	\label{eq:momentum quantum-ring Hamiltonian}
\end{equation}
where $c^{\dagger}_{k}$ and $c_{k}$ are the creation and annihilation operators for the states in the quantum-ring, $k = 2\pi m/N$ with $m = 0,1,2,\dots N-1$. In Majorana representation, the Hamiltonian is written as:
\begin{equation}
	H_{\text{R}} = i2t \sum_{k}\cos\left(k + \frac{2\pi}{N}\Phi\right)\alpha^{\left(1\right)}_{k}\alpha^{\left(2\right)}_{k}\,,
\end{equation}
with $c_{k} = (\alpha^{\left(1\right)}_{k} + i\alpha^{\left(2\right)}_{k})/\sqrt{2}$ and $c^{\dagger}_{k} = (\alpha^{\left(1\right)}_{k} - i\alpha^{\left(2\right)}_{k})/\sqrt{2}$. The last Hamiltonian corresponds to the coupling between the fermionic states of the TSCs and the fermionic states in the quantum-ring with position $l = 0$ and $l = j$ being $j = 1,2, \dots , N - 1$. This Hamiltonian in real space reads:
\begin{equation}\label{H5}
	H_{\text{C}} = \lambda f^{\dagger}c_{0} + \lambda' e^{i\frac{\theta}{2}}a^{\dagger}c_{j} + \text{h.c.}\,.
\end{equation}
Here, $f^{\left(\dagger\right)}$ are the fermionic states of one TSC coupled to the site $l = 0$ in the quantum-ring and $a^{\left(\dagger\right)}$ represents the fermionic states of the TSC coupled to the site $l = j$ with amplitudes $\lambda$ and $\lambda'$, respectively, and with $\theta$ as the superconducting phase difference. Therefore, a low energy coupling Hamiltonian can be obtained by introducing the transformation $f^{\dagger} = (\mu \eta_{1} - i\nu \eta_{2})/\sqrt{2}$ and $a^{\dagger} = (\delta \eta_{3} - i\kappa \eta_{4})/\sqrt{2}$, with $\left|\mu\right|^{2} + \left|\nu\right|^{2} = \left|\delta\right|^{2} + \left|\kappa\right|^{2} =1$. The parameters $\mu$, $\nu$, $\delta$ and $\kappa$ are the weights of the quasiparticle excitations. As we are working in the momentum space, the transformation $c_{0} = \sum_{k}c_{k}$ and $c_{j} = \sum_{k}e^{ikj}c_{k}$ must be taken into account:

\begin{align}
\begin{split}
H_{C} = &i\sum_{k}\left[\lambda_{1}\eta_{1}\alpha^{\left(2\right)}_{k} + \lambda_{2}\alpha^{\left(1\right)}_{k}\eta_{2}\right] \\
+ &i\lambda'_{1}\sum_{k}\left[S_{j}\left(k,\theta\right)\alpha^{\left(1\right)}_{k}\eta_{3} + C_{j}\left(k,\theta\right)\eta_{3}\alpha^{\left(2\right)}_{k}\right]\\
+&i\lambda'_{2}\sum_{k}\left[C_{j}\left(k,\theta\right)\alpha^{\left(1\right)}_{k}\eta_{4} + S_{j}\left(k,\theta\right)\eta_{4}\alpha^{\left(2\right)}_{k}\right]\,,
\label{eq:Coupling Hamiltonian}
\end{split}
\end{align} 
where $\lambda_{1} = \mu\lambda$, $\lambda_{2} = \nu\lambda$ and $\lambda'_{1} = \delta\lambda'$, $\lambda'_{2} = \kappa\lambda'$. Note if $\nu \rightarrow 0$ and $\kappa \rightarrow 0$, the Hamiltonian reduces to the case in which the two nanowires are infinite and a truly Majorana zero mode hybridizes with the quantum-ring. In the equation above $S_{j}\left(k,\theta\right) = \sin\left(kj +\theta/2\right)$ and $C_{j}\left(k,\theta\right) = \cos\left(kj +\theta/2\right)$. Hence, the full Hamiltonian of the model is written as:
\begin{equation}
H = H_{\text{R}} + H_{\text{TSC}} + H_{\text{C}}\,.
\end{equation}

Figure\ \ref{fig:schematic_model} displays a schematic representation of the system for $H_{\text{C}}$ given by Eq.\ (\ref{H5}) [Figure \ref{fig:schematic_model}($a$)] and Eq.\ (\ref{eq:Coupling Hamiltonian}) [Figure \ref{fig:schematic_model}($b$)].

\subsection{Bogoliubov-de Gennes}

To obtain the energy spectra, we write the Hamiltonian in a BdG form, which implies writing it in a redundant manner. This procedure, apart from returning a skew-symmetric matrix in Majorana representation, imposes a constraint which is usually called particle-hole symmetry. Therefore, all the spectra obtained will be symmetric around zero energy. The basis chosen for the skew-symmetric matrix is
\begin{equation*}
\Psi = \left(\alpha^{\left(1\right)}_{0}\alpha^{\left(1\right)}_{1}\,\dots\,\alpha^{\left(1\right)}_{N-1}\alpha^{\left(2\right)}_{0}\alpha^{\left(2\right)}_{1}\,\dots\,\alpha^{\left(2\right)}_{N-1} \eta_{1}\eta_{2}\eta_{3}\eta_{4}\right)^{T}. 
\end{equation*}
Hence, the Hamiltonian is written as: 
\begin{equation}
	H =\Psi^{\dagger} H_{\text{BdG}} \Psi. 
\end{equation} 
The size of $H_{\text{BdG}}$ depends on the number of states in the quantum ring. Here, we can identify the symmetry of the Hamiltonian by observing on which basis the Hamiltonian can be written as a linear combination of the generators of the symmetry group. For example, if we consider a single mode in the quantum-ring, $H_{\text{BdG}}$ has a dimension of $6 \times 6$ only if $\xi_{1\left(2\right)} \neq 0$ and can be written as a linear combination of the imaginary elements of the $SU\left(6\right)$ in the $\lambda$-representation. Thus, if we consider $N$ the number of states in the quantum-ring, the symmetry of the Hamiltonian is $SU\left(2N + 4\right)$ in Majorana representation \cite{schnyder2008classification}. Therefore, the Hamiltonian can be written in compact form as:

\begin{equation}
	H_{\text{BdG}} = \frac{1}{2}\mathbf{h}\cdot\mathbf{\Lambda}\,,	
\end{equation}
with $\mathbf{h}$ a vector with the dimension $\left(2N + 4\right)^{2} -1$, which is nothing but the dimension of the $SU\left(2N + 4\right)$ Lie group and $\mathbf{\Lambda}$ are the elements of the Lie group.
\subsection{Chiral Symmetry}
We have imposed over the Hamiltonian a $\mathcal{PH}$ constraint. Therefore, the eigenvalues of the $H_{\text{BdG}}$ come in pairs. What remains is to determine at which points in the parameter space the system possesses $\mathcal{C}$ symmetry. Hence, if the Hamiltonian preserves $\mathcal{C}$ symmetry it must satisfy the following property  
\begin{equation}
\mathbf{\Gamma} H_{\text{BdG}} \mathbf{\Gamma}^{-1} = -H_{\text{BdG}}\,,
\label{eq:chiral}   
\end{equation}
with $\Gamma$ being a chiral operator written in a general form as
\begin{equation}
\mathbf{\Gamma} = e^{-i\gamma\mathbf{n}\cdot\mathbf{\Lambda}}\,,
\end{equation}
where, $\gamma \mathbf{n}$ are $\left(2N + 4\right)^{2} -1$ arbitrary parameters. Hence, the rotation is made around the unit vector $\mathbf{n}$. Choosing $\gamma = \pi$ and $\mathbf{n} \cdot \mathbf{h} = 0$, the operator satisfies Eq.\ (\ref{eq:chiral}) and the anticommutation relation $\left\{\mathbf{\Gamma},H_{\text{BdG}}\right\} = 0$. As $\mathbf{\Gamma}^{2} = 1$, the eigenvalues of the operator are $\mathbf{\Gamma} = \pm 1$. Following Ref. \cite{sato2011topology}, we know two things: (i) The zero energy crossing states in a given BdG Hamiltonian are simultaneously eigenstates of the $\mathbf{\Gamma}$ operator; and (ii) nonzero energy states are described by a linear combination of two states $\psi_{\pm}$ precisely associated with the eigenvalues of $\mathbf{\Gamma}$. Therefore, for an eigenstate outside zero energy $\psi = c_{+}\psi_{+} + c_{-}\psi_{-}$ with $\left|c_{+}\right| = \left|c_{-}\right|$.

\section{Junction Configurations and Spectra} \label{Sec.3}
\subsection{Junctions}

\begin{figure*}[ht]
	\centering
	\includegraphics[scale = 1]{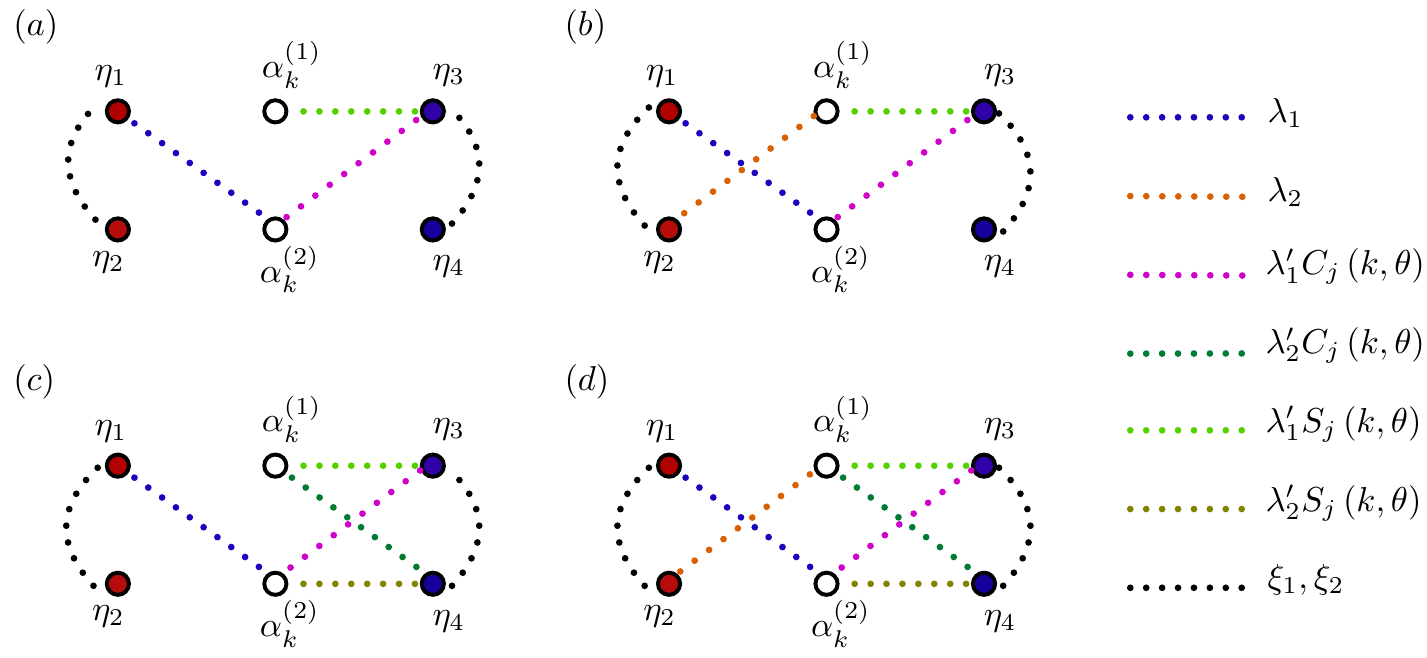}
	\caption{\label{fig:Four_couplings} Schematic configurations of the four coupling configurations. $\left(a\right)$ A-junction: two MBSs are hybridized directly with the quantum-ring. $\left(b\right)$ B-junction: three MBSs hybridized directly. $\left(c\right)$ C-junction: three MBSs hybridized directly, where two of the MBSs belong to the TSC with phase $\theta$. $\left(d\right)$ D-junction: four MBSs hybridized directly with the quantum-ring. In all the junctions we have the possibility of indirect coupling between $\eta_{2}$ and $\eta_{4}$, through $\xi_{1\left(2\right)}$. The asymmetry in the couplings is due to the presence of the phase difference $\theta$.}
\end{figure*}

Before proceeding with the spectra, we consider four types of junction configurations for which the minimum number of MBSs hybridizing the ring is two, and the maximum number is four. For the latter case, when $\xi_{1\left(2\right)}=0$, we have a pretty similar configuration as that of Ref.\ \cite{houzet2019majorana}, emphasizing
that we have a single phase difference. Figure \ref{fig:Four_couplings} shows the graphical representation for the four types of junctions considered. This is a graphical representation of the coupling Hamiltonian. Note that there is a coupling between MBSs with the same chirality, like $\eta_{3}$ and $\alpha^{\left(1\right)}_{k}$. This is purely due to the effect of superconducting phase difference, and it can be checked by taking $\theta = 0$ in Eq.\ (\ref{eq:Coupling Hamiltonian}), returning to the symmetric coupling. In these four junctions, we consider two possibilities: a direct and an indirect hybridization of the MBSs. Such considerations give rise to effects in the delocalization of the MBSs \cite{medina2020influence,deng2018nonlocality}.

In addition to the phase difference, we have taken into account the position of the TSCs, since that starts to be important for $N \geq 4$. For $N = 3$, no matter what the TSCs positions are, the system is non-centrosymmetric, while in the case of $N = 4$, the system could be either: non-centrosymmetric or centrosymmetric, for $j = 1$ and $j = 2$, respectively.

\subsection{Spectra}

In this section, we are going to analyze the spectra for each of the junctions mentioned before. In what follows, the energy spectra are presented in units of twice the hopping between states in the quantum-ring, i.e. in units of $2t$. We focus only on a system with $N = 3$ and $N = 4$ number of sites in the ring. With this restriction, we lost some generality, but not the utility of the results, since some Aharonov-Bohm quantum-rings architecture is based on InAs, for example, which could be used to prove the results \cite{whiticar2020coherent,potts2019electrical}. 

\subsubsection{Quantum-ring with three quantum dots.}
\begin{figure*}[ht]
	\centering
	\includegraphics[scale=0.5]{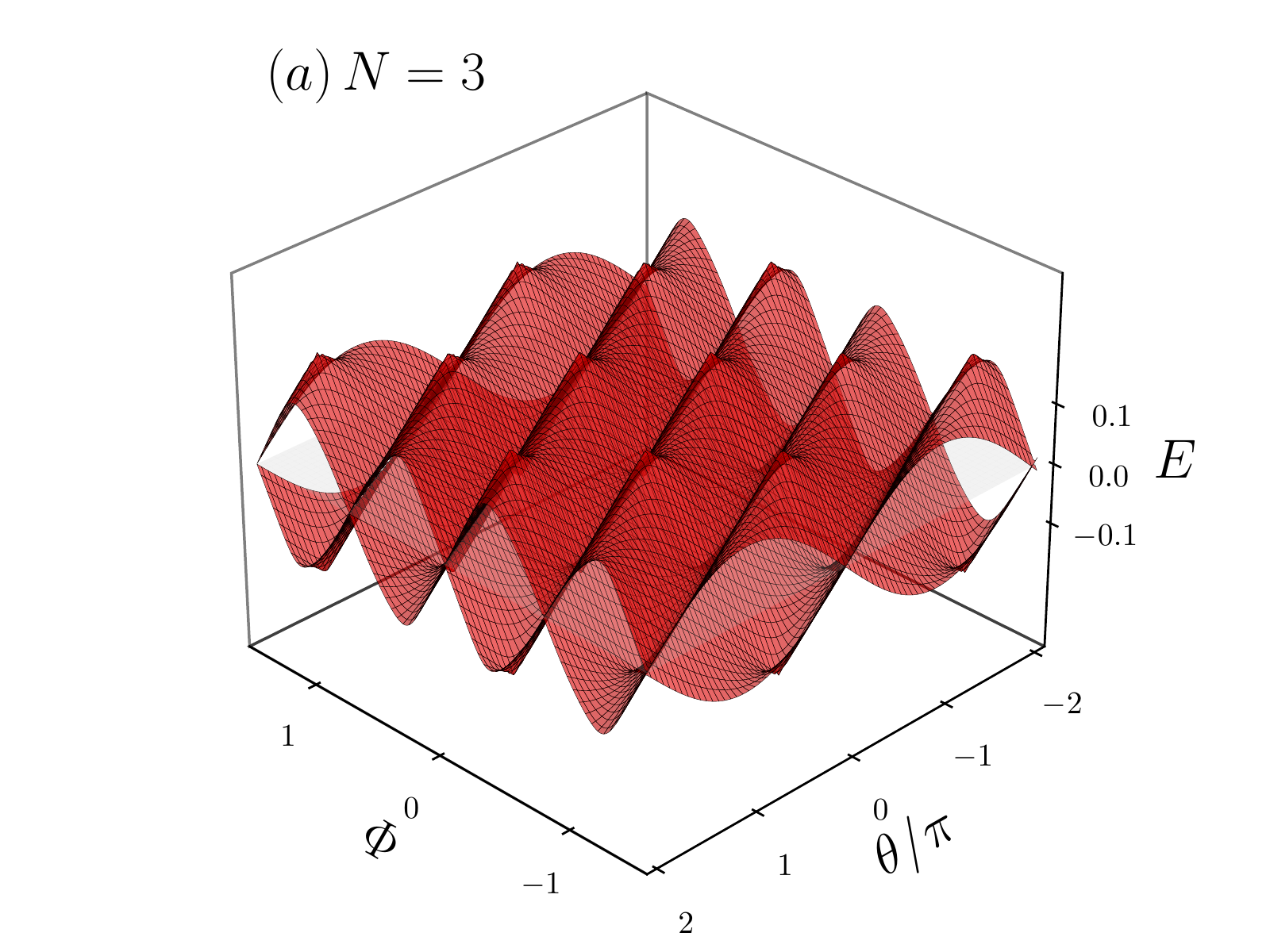}\includegraphics[scale=0.5]{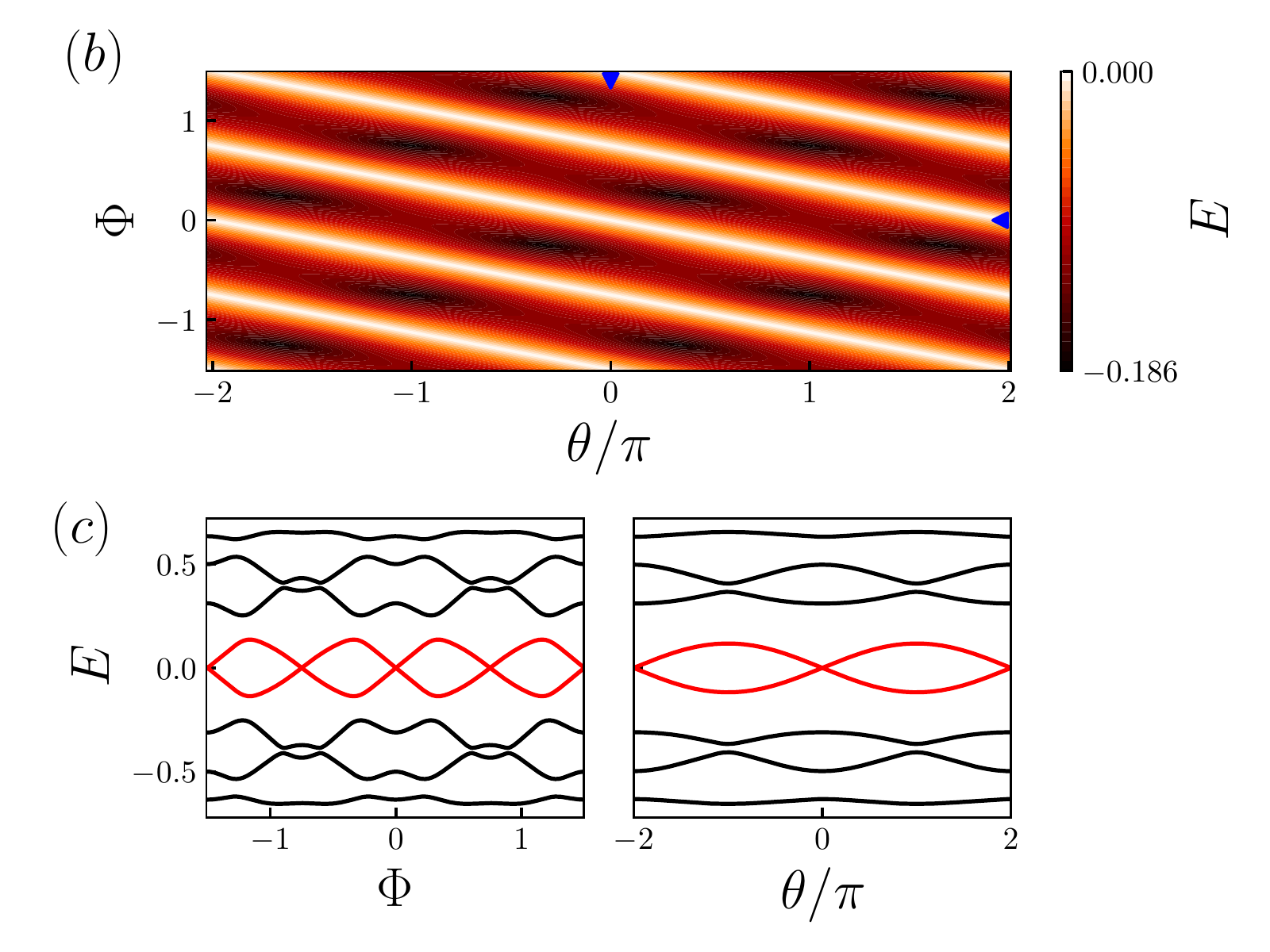}
	\caption{\label{fig:spec1} Quantum-ring spectra in the $A$-junction configuration using $N = 3$, for coupling amplitudes $\lambda_{1} = \lambda'_{1} = 0.5$ and $\lambda_{2} = \lambda'_{2} = \xi_{1} = \xi_{2} = 0$. $(a)$ MBSs subgap bands in the parameter space. $(b)$ Color map of the negative energy subgap band projection, showing the topological pattern of line nodes. $(c)$ 2D representation of the full spectra for fixed values $\theta = 0$ (left panel) and $\Phi = 0$ (right panel). Blue triangles in $(b)$ represent the fixed values of flux and phase difference at which the 2D spectra is plotted in $(c)$.}
\end{figure*}
\begin{figure*}[ht]
	\centering
	\includegraphics[scale=0.5]{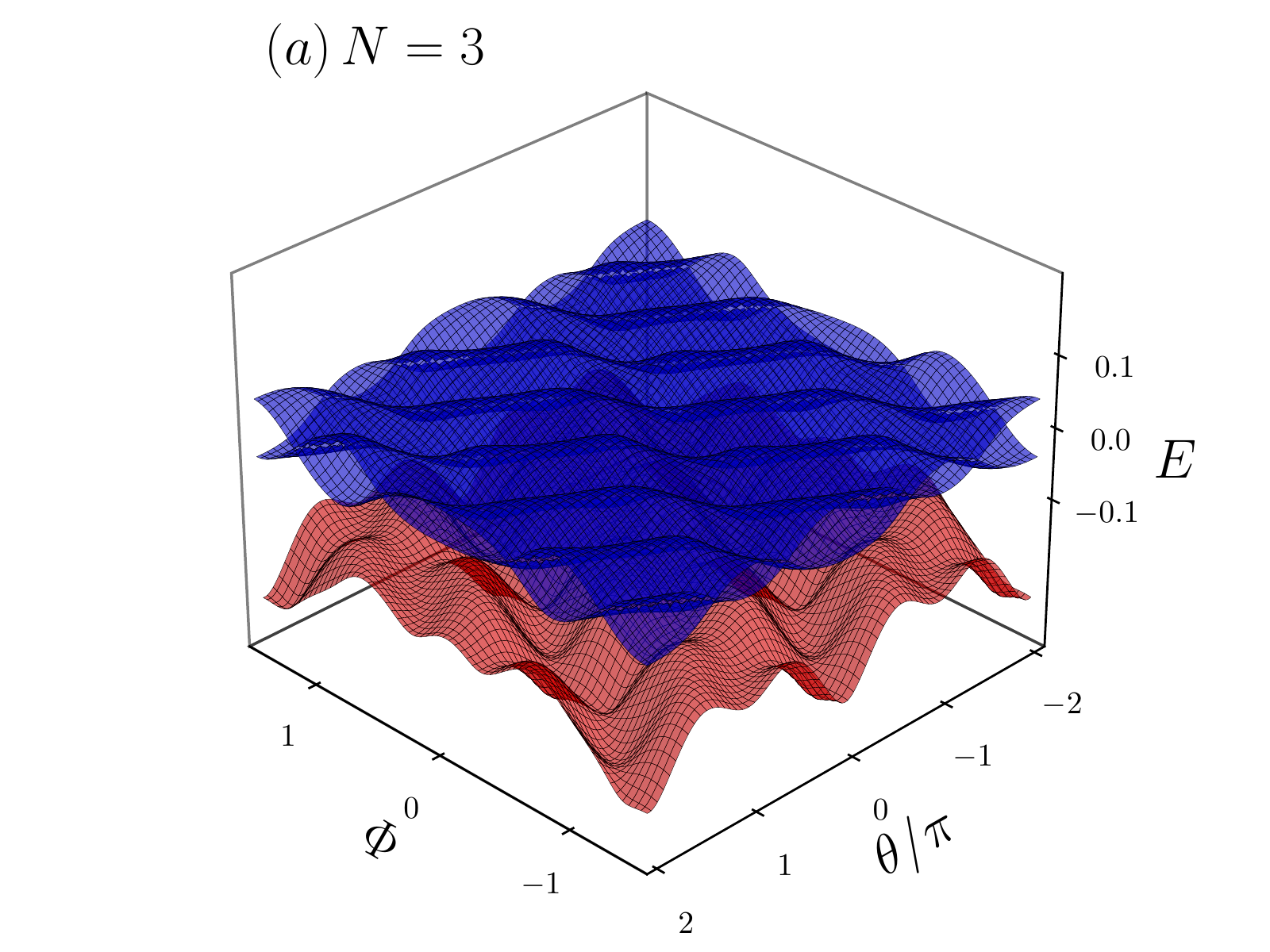}\includegraphics[scale=0.5]{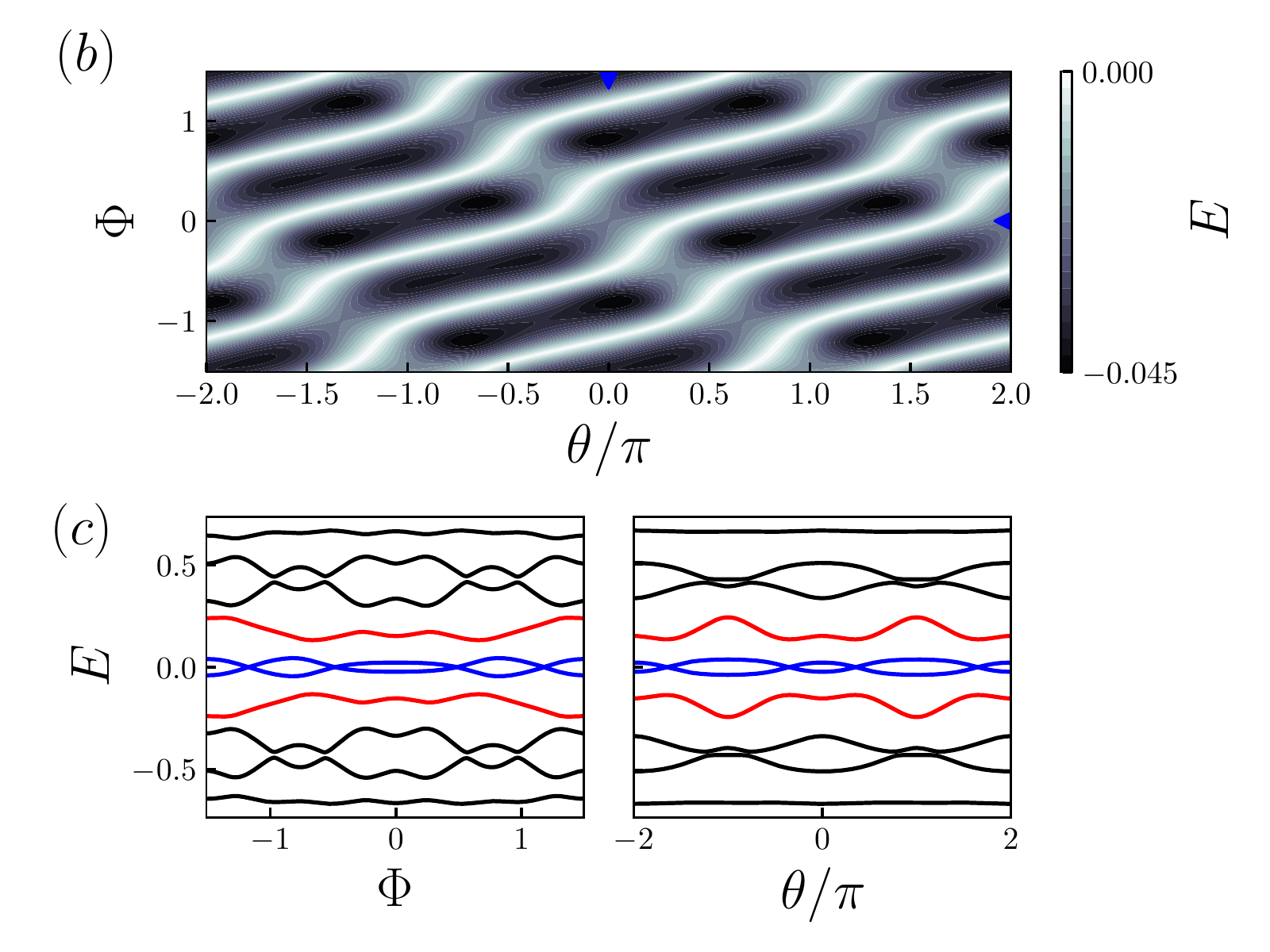}
	\caption{\label{fig:spec4} Quantum-ring spectra in $C$-junction configuration with $N = 3$  and coupling amplitudes $\lambda_{1} = \lambda'_{1} = 0.5$, $\lambda_{2} =  0$, $\lambda'_{2} = 0.3$ and $\xi_{1} = \xi_{2} = 0.2$. $(a)$ MBSs subgap bands in the parameter space. $(b)$ Color map of the negative energy subgap band projection, showing the undulated nodes topological pattern. $(c)$ 2D representation of the full spectra for fixed values $\theta = 0$ (left panel) and $\Phi = 0$ (right panel). Blue triangle marks in $(b)$ represent the fixed values of flux and phase difference at which the 2D spectra is plotted in $(c)$.}
\end{figure*}
As mentioned before, we considered four types of junctions named: A-junction, B-junction, C-junction, and D-junction. In the A-junction, only two MBSs hybridize the system and their spectra can be observed in Figure \ref{fig:spec1}. The spectra are displayed in the parameter space formed by the magnetic flux, $\Phi$, and the phase difference between both TSCs, $\theta$, in a non-centrosymmetric configuration with coupling parameters $\lambda_{2} = \lambda'_{2} = \xi_{1} = \xi_{2} = 0$. Note that for this case ($N=3$) only the non-centrosymmetric configuration is allowed. The panel $(a)$ shows the inner gap bands formed by the hybridized MBSs, and from this we can see that bands have zero energy crossings. Panel $(b)$ shows the projection of the lower band utilizing a color map. It is clear from the plots that linear node patterns are formed, having a periodicity of $4\pi$ concerning the phase difference $\theta$. In addition, panel $(c)$ shows the 2D full spectra for $\Phi = 0$ and $\theta = 0$.
\begin{figure*}[ht]
	\centering
	\includegraphics[scale=0.5]{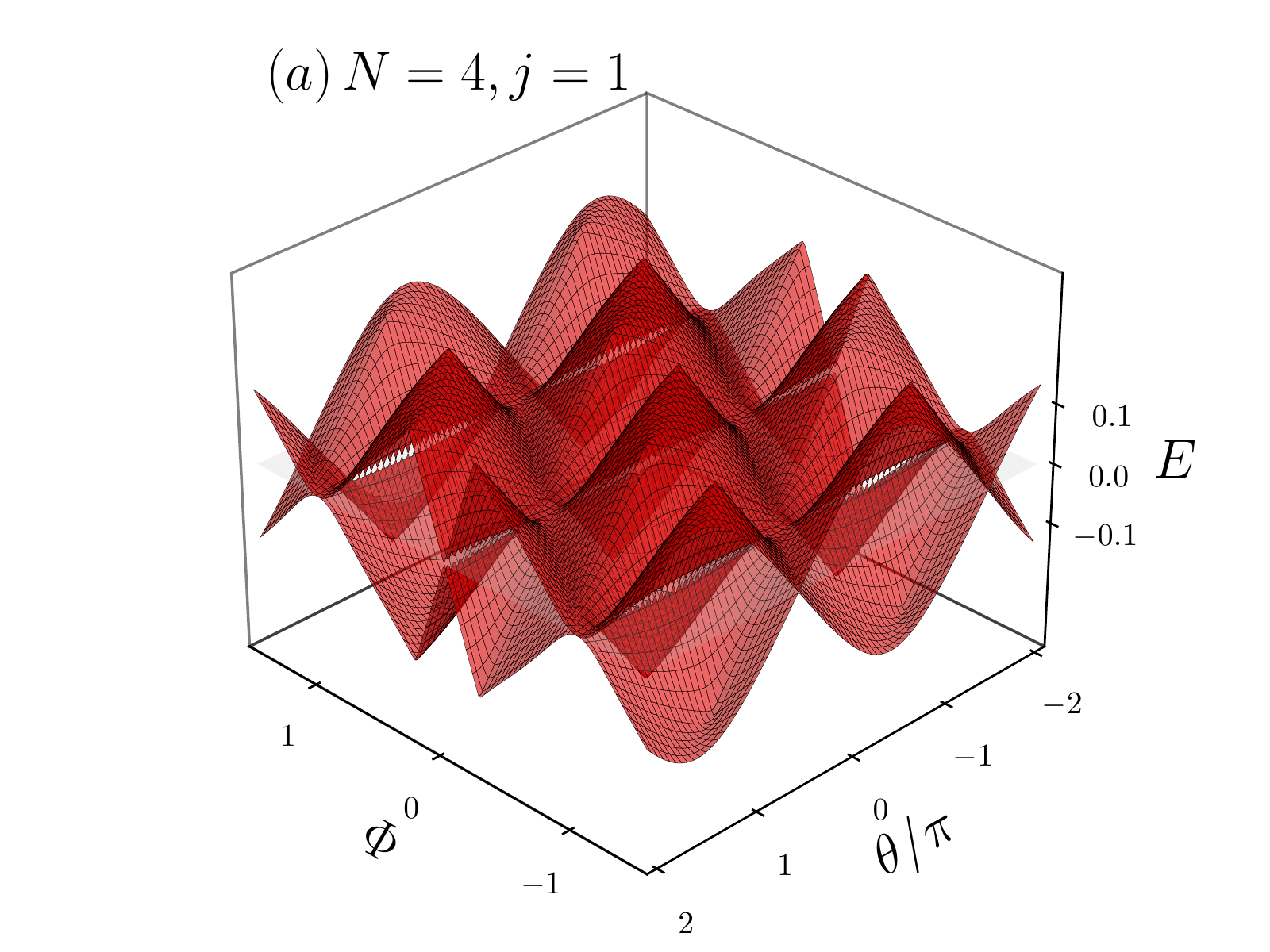}\includegraphics[scale=0.5]{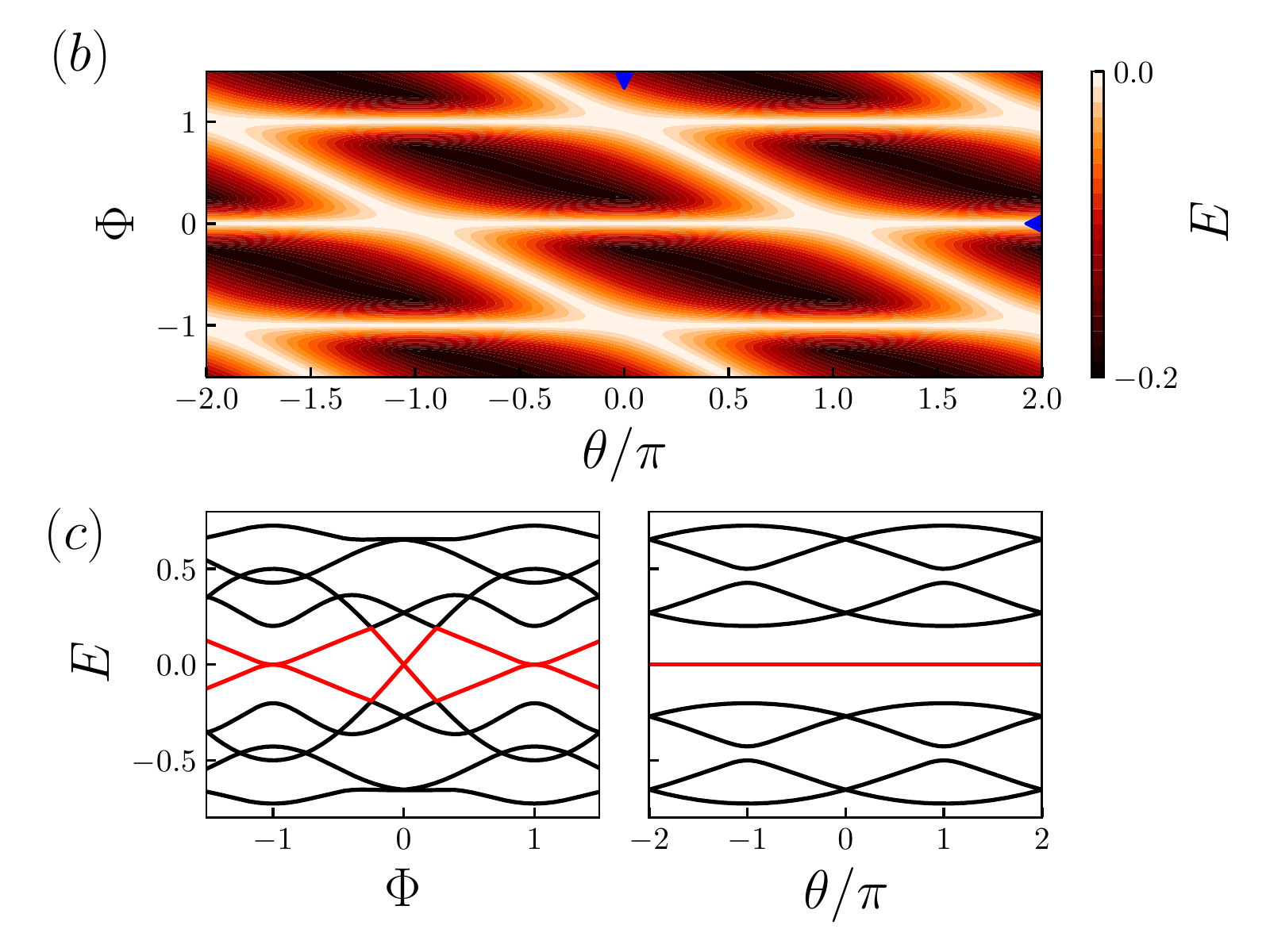}
	\caption{\label{fig:spec2} Quantum-ring spectra in the $A$-junction configuration with $N = 4, j = 1$ with couplings $\lambda_{1} = \lambda'_{1} = 0.5, \lambda_{2} = \lambda'_{2} = 0$ and $\xi_{1} = \xi_{2} = 0$.  $(a)$ MBSs subgap bands in the parameter space. $(b)$ Color map of the negative energy subgap band projection, showing rectangular nodes topological pattern. $(c)$ 2D representation of the full spectra for fixed values $\theta = 0$ (left panel) and $\Phi = 0$ (right panel). Blue triangle marks in $(b)$ represent the fixed values of flux and phase difference at which the 2D spectra is plotted in $(c)$.}
\end{figure*}
 As observed in Figure \ref{fig:Four_couplings}, all junctions may have interactions between MBSs from the same TSC. Therefore, in A-junction we have three additional parameter configurations obtained: when $\xi_{1} = 0$ and $\xi_{2} = 0.2$, the reverse one $\xi_{1} = 0.2$ and $\xi_{2} = 0$, and finally when $\xi_{1} = \xi_{1} = 0.2$. The former two cases can be understood by considering the length of one TSC to be finite, while the other two nanowires are finite TSCs. One finite TSC generates the overlapping of the MBSs from the same TSC, removing the MBSs from zero energy and two bands that form a subgap in the spectra. The remaining Majorana has no other path that forms a flat band at zero energy. When two finite TSCs are considered, the spectra display four bands inside the gap, with two lower energy bands forming line nodes. Therefore, under those couplings with A-junction configuration, we obtain line and flat band. In this paper, we focus on the case with zero energy crossings and on the formation of topologically protected patterns (Figure \ref{fig:Four_couplings}), therefore the latter cases using $\xi_{2(1)}\neq\xi_{1(2)}=0$ and $\xi_{1}=\xi_{2}\neq 0$ are not presented in figures.
 
\begin{figure*}[ht]
	\centering
	\includegraphics[scale=0.5]{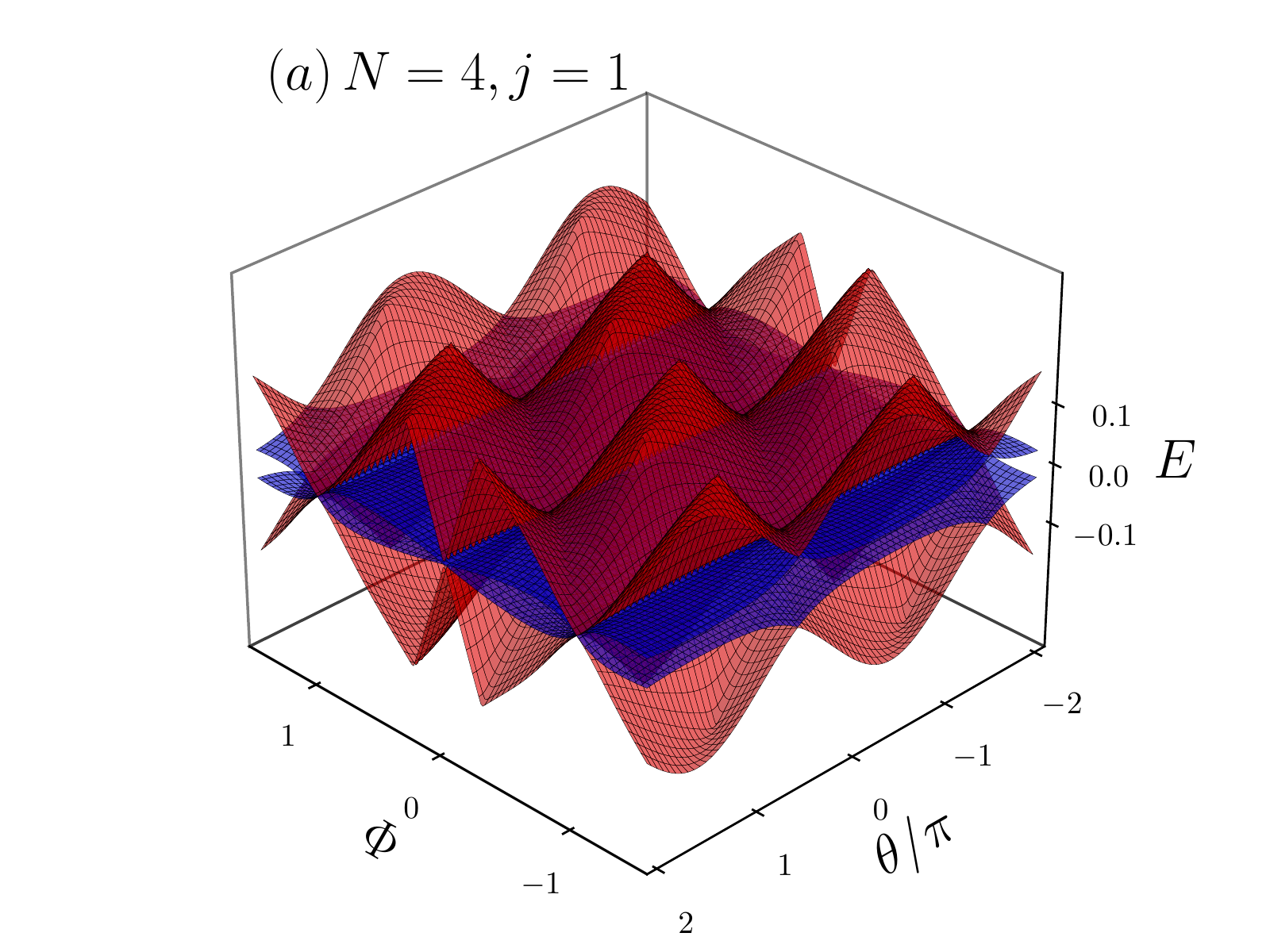}\includegraphics[scale=0.5]{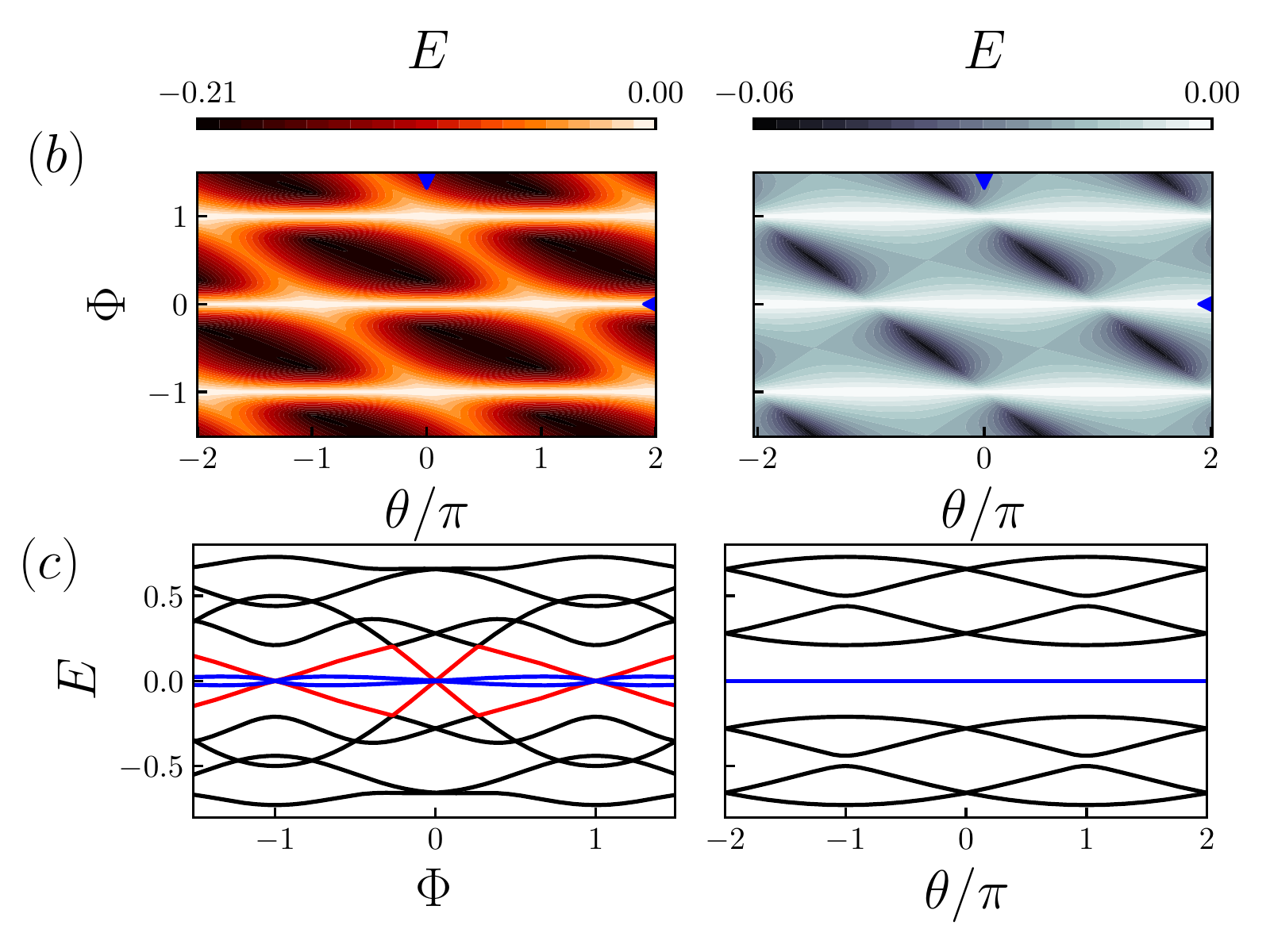}
	\caption{\label{fig:spec3} Quantum-ring spectra in $A$-junction configuration with $N = 4, j = 1$, the couplings are $\lambda_{1} = \lambda'_{1} = 0.5$, $\lambda_{2}=\lambda'_{2} = 0$ and $\xi_{1} = \xi_{2} = 0.2$. $(a)$ MBSs subgap bands in the parameter space. $(b)$ Color map of the negative energy subgap band projection, showing four-fold line nodes topological pattern.  Left and right panels correspond to red and blue bands in panel $(c)$, respectively. $(c)$ 2D representation of the full spectra for fixed values $\theta = 0$ (left panel) and $\Phi = 0$ (right panel). Blue triangle marks in $(b)$ represent the fixed values of flux and phase difference at which the 2D spectra is plotted in $(c)$.}
\end{figure*}
\begin{figure*}[h!]
	\centering
	\includegraphics[scale=0.5]{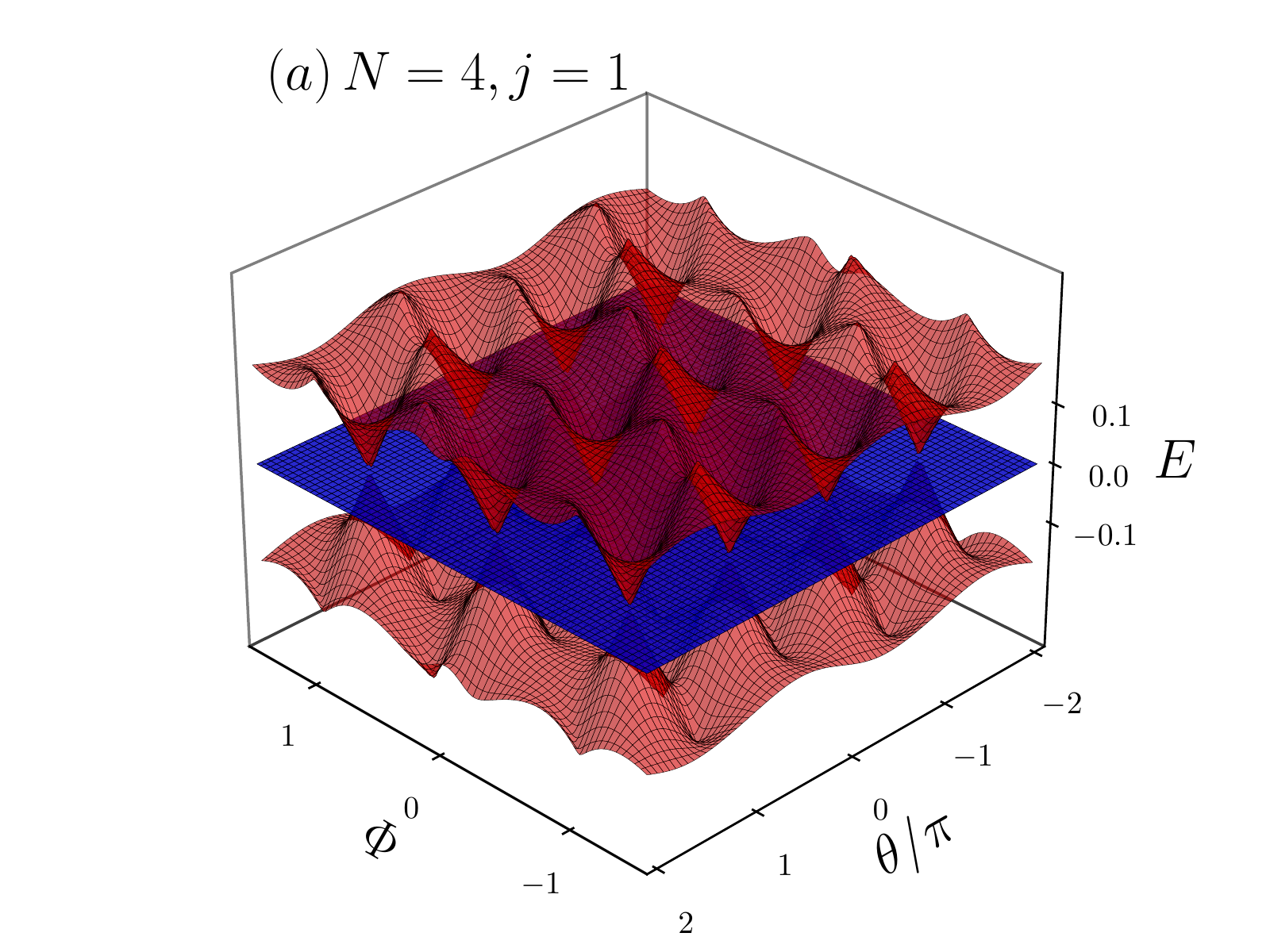}\includegraphics[scale=0.5]{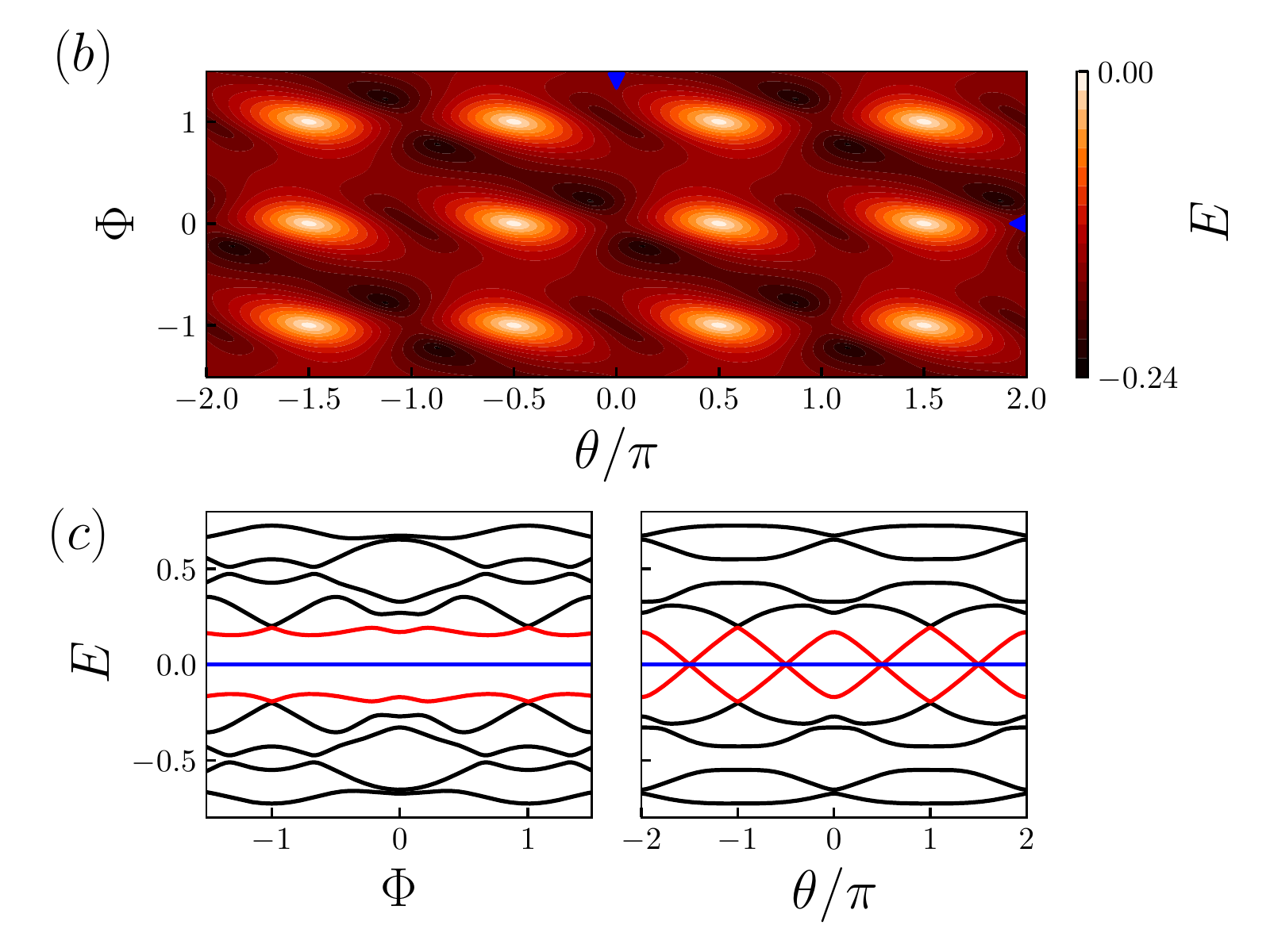}
	\caption{\label{fig:spec5} Quantum-ring spectra in $C$-junction configuration with $N = 4, j = 1$ and coupling amplitudes $\lambda_{1} = \lambda'_{1} = 0.5$, $\lambda_{2} =  0$, $\lambda'_{2} = 0.3$ and $\xi_{1} = \xi_{2} = 0$. $(a)$ MBSs subgap bands in the parameter space. $(b)$ Color map of the negative energy subgap band projection, showing point nodes topological pattern. $(c)$ 2D representation of the full spectra for fixed values $\theta = 0$ (left panel) and $\Phi = 0$ (right panel). Blue triangle marks in $(b)$ represent the fixed values of flux and phase difference at which the 2D spectra is plotted in $(c)$.}
\end{figure*}
\begin{figure*}[h!]
	\centering
	\includegraphics[scale=0.5]{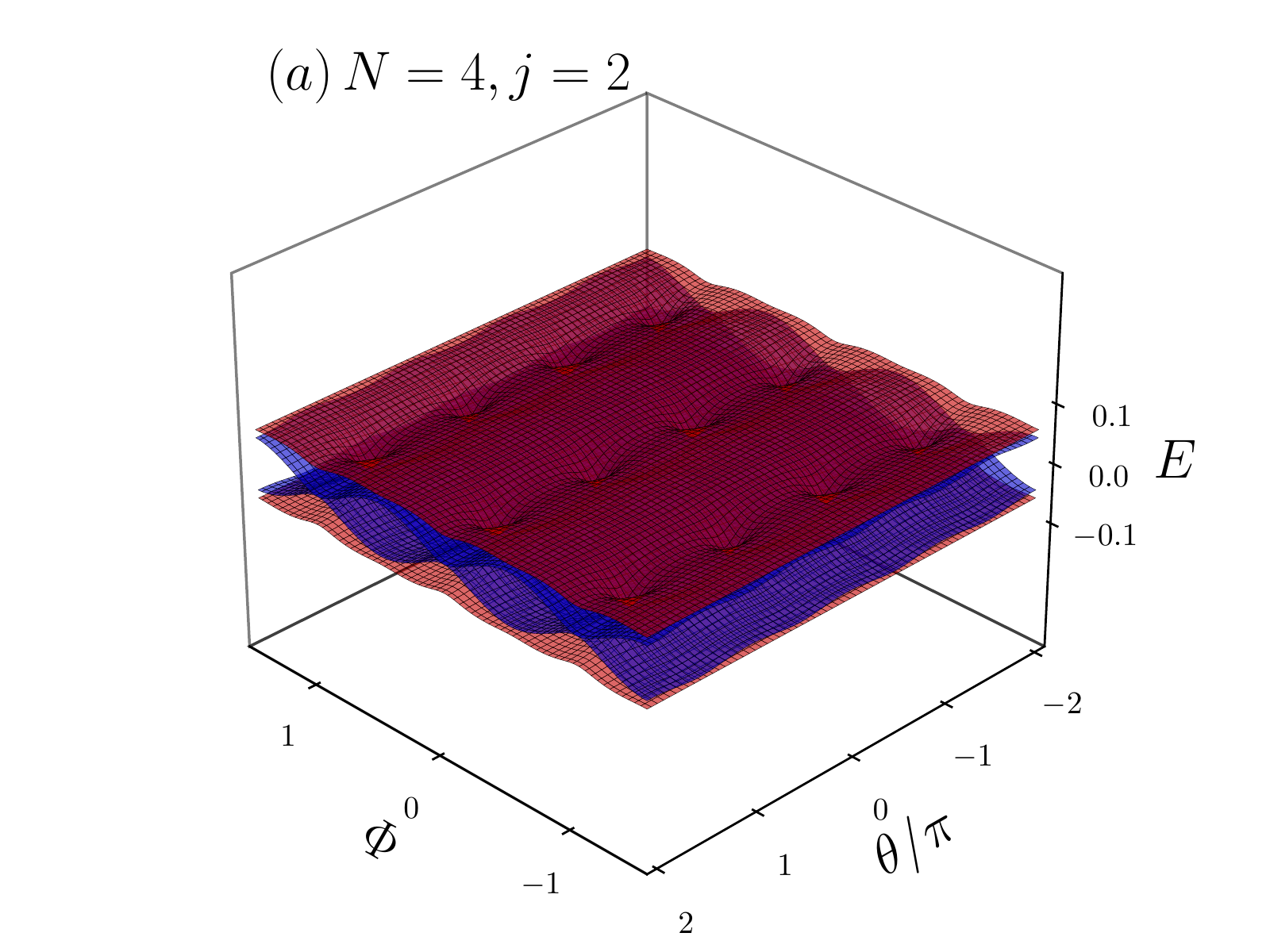}\includegraphics[scale=0.5]{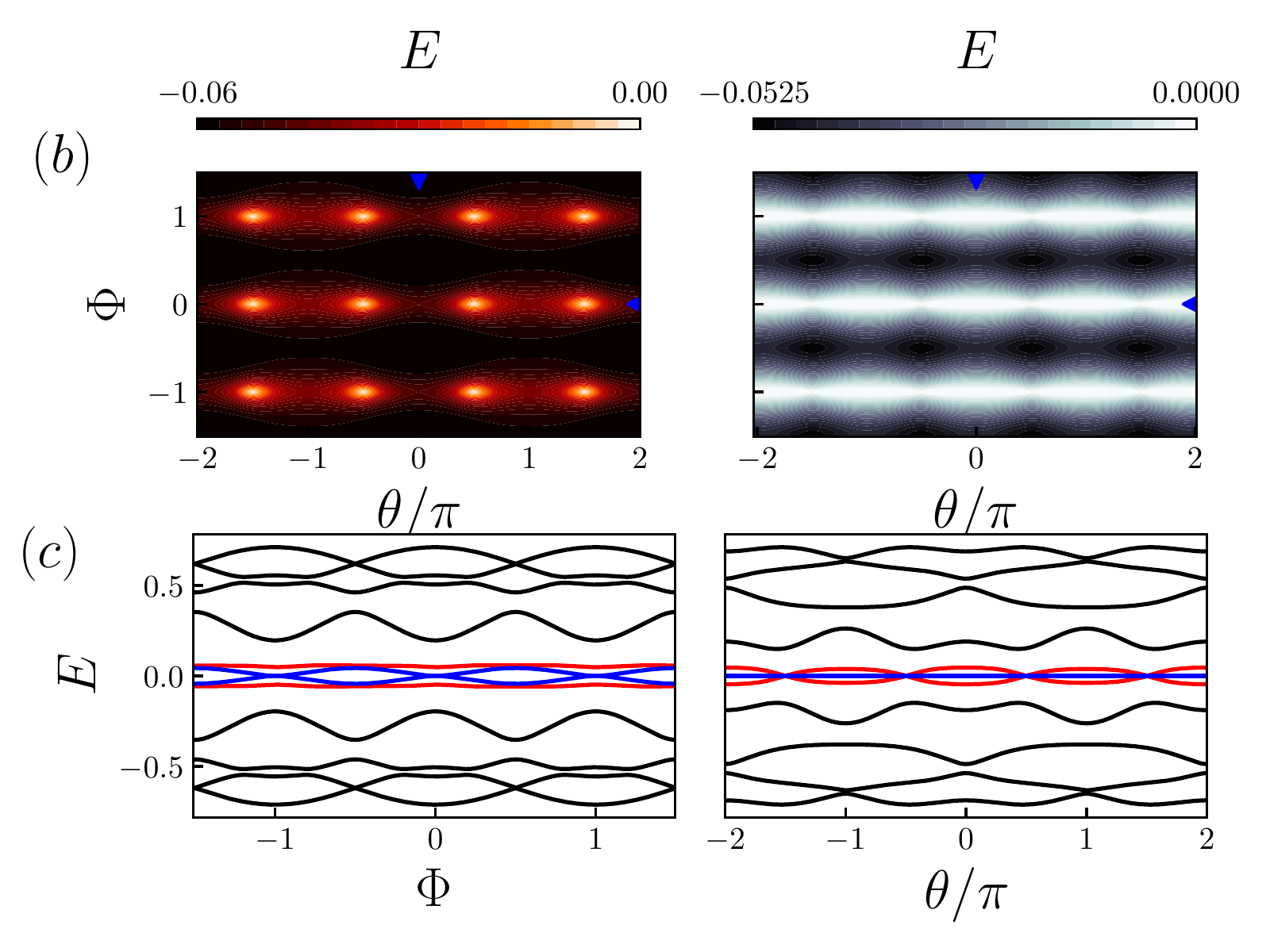}
	\caption{\label{fig:spec6}Quantum-ring spectra in $C$-junction configuration with $N = 4, j = 2$ and coupling amplitudes $\lambda_{1} = \lambda'_{1} = 0.5$, $\lambda_{2} =  0$, $\lambda'_{2} = 0.3$ and $\xi_{1} = \xi_{2} = 0.2$. $(a)$ MBSs subgap bands in the parameter space. $(b)$ Contour plot representing the projection of the negative energy subgap band, showing point nodes topological pattern. $(c)$ 2D representation of the full spectra for fixed values $\theta = 0$ (left panel) and $\Phi = 0$ (right panel). Blue triangle marks in $(b)$ represent the fixed values of flux and phase difference at which the 2D spectra is plotted in $(c)$.}
\end{figure*}

In the B-junction configuration, the system presents only a flat band for any of the couplings as long as $\xi_{2} = \lambda'_{2} = 0$. In contrast with the B-junction, the C-junction allows tuning of $\xi_{1}$ and $\xi_{2}$. Therefore, when this junction is considered, it is possible to obtain line nodes, undulated nodes, and flat bands. We do not graphically present all these behaviors mentioned since line nodes are shown already and an example of a flat band will be shown below in another case. However, undulated nodes can be observed in Figure \ref{fig:spec4}, where the spectra are presented for C-junction configuration using the parameters $\lambda_{1} = \lambda'_{1} = 0.5$, $\lambda_{2} = 0$, $\lambda'_{2} = 0.3$ and $\xi_{1} = \xi_{2} = 0.2$. Panel $(a)$ shows the MBSs bands, and panel (b) shows a color map of the negative energy. Two hybridized Majorana bands are observed in these figures, which are better observed in panel (c) in red and blue curves. The red band, with a particle-hole symmetric upper band, forms a subgap in the spectra while the blue band, inside the subgap, forms a zero energy crossing, and in the parameter space, it forms undulated nodes patterns, which correspond to the deformation of the line nodes. 

Finally, we address the D-junction configuration, which consists of the direct hybridization of the four MBSs in the quantum-ring. For this junction configuration, line nodes and undulated nodes are obtained, being similar to those presented in Figures. \ref{fig:spec1} and \ref{fig:spec4}. We do not graphically present such results. Henceforth, all the possible couplings for the four types of junctions that display zero energy crossing and therefore regions where $\mathcal{C}$ symmetry is preserved are summarized in Table \ref{table1}.  

\subsubsection{Quantum-ring with four quantum dots.}

As in the previous subsection, we are not graphically presenting all and each of the nodes observed, since many of them repeatedly appear in the different configurations considered.

For $N =4$, the system possesses two different scenarios: centrosymmetric or non-centrosymmetric ($j = 2$ and $j = 1$, respectively). In this sense, there is a richness in its topological characteristics, since different topological patterns appear according to this configuration. For A-junction, in a non-centrosymmetric case, the system displays three types of nodes: rectangular formation of nodes when $\xi_{1} = \xi_{2} = 0$; line nodes coexisting with a flat band for $\xi_{1} = 0.2$ and $\xi_{2} = 0$ (not shown graphically); and a four-fold line node when four MBSs display the same line nodes simultaneously for $\xi_{1} = \xi_{2} = 0.2$. In Figure \ref{fig:spec2} ($\xi_{1} = \xi_{2} = 0$), there is a graphical representation of the formation of rectangular nodes. The rectangular node pattern is formed by two line nodes coming from different directions in the parameter space: one is formed in $\mathcal{T}$ reversal points for a ring with $N = 4$ sites ($\Phi = \dots,-1,0,1,\dots$) and remain constant regardless $\theta$, and the second line nodes depend on $\theta$ and $\Phi$. On the other hand, in Figure \ref{fig:spec3} we present the spectra using $\xi_{1} = \xi_{2} = 0.2$, from which four-fold line nodes are observed. Note that in this case, two Majorana are coupled directly, and the other two are coupled indirectly. For this case there is no formation of a subgap, but rather the four MBSs oscillate around zero energy, sharing the same zero energy points, as can be observed in panel $\left(b\right)$ and $\left(c\right)$. In a B-junction, line nodes are displayed when $\xi_{1} = \xi_{2} = 0.2$ and flat bands when $\xi_{1} = 0.2$ and $\xi_{2} = 0$ (not shown graphically). Figure \ref{fig:spec5} shows the spectra for the C-Junction configuration with a non-centrosymmetric arrangement, where three MBSs are coupled directly and the TSCs are infinitely long ($\xi_{1} = \xi_{2} = 0$). The spectra show the coexistence of point nodes and flat bands. The point nodes are formed by the red bands shown in the panel $\left(a\right)$ and are located at periodic points in phase difference $\theta$, and along the $\mathcal{T}$ reversal symmetry points in $\Phi$. Since we are considering a spinless or a fully polarized system, these topological point nodes in the parameter space can be seen as Majorana-Weyl cones, which are non-degenerate bands. There are some circuits that can simulate Weyl points without TSCs \cite{fatemi2021weyl} and with TSCs \cite{kotetes2019synthetic}. By keeping the same coupling and moving one of the TSCs from site $j = 1$ to site $j = 2$, the system results in a centrosymmetric system, and the point nodes are robust throughout this change, as can be seen in Figure \ref{fig:spec6}. The difference is shown with the formation of line nodes coexisting with the point nodes: a pair of MBSs are forming the point nodes while the other two form a line node precisely at $\mathcal{T}$ reversal symmetric points of a ring with $N = 4$.

Finally, in the D-junction configuration, where all MBSs are hybridized directly in the ring and with non-centrosymmetric arrangement, the spectra only display line nodes, while in the case of the centrosymmetric system we can have point and points coexisting with line nodes. Those results are not shown graphically, since they are already shown for other types of configurations. All the results for a ring formed with four quantum-dots are summarized in Table \ref{table2}. All possible couplings that are not included in the table break $\mathcal{C}$ symmetry, and as a consequence topological patterns are not formed. \cite{teo2010topological, beenakker2013fermion}. 

\section{Persistent and dc Josephson currents}\label{Sec.4}

\begin{figure*}[tbph]
	\centering
	\includegraphics[scale=0.45]{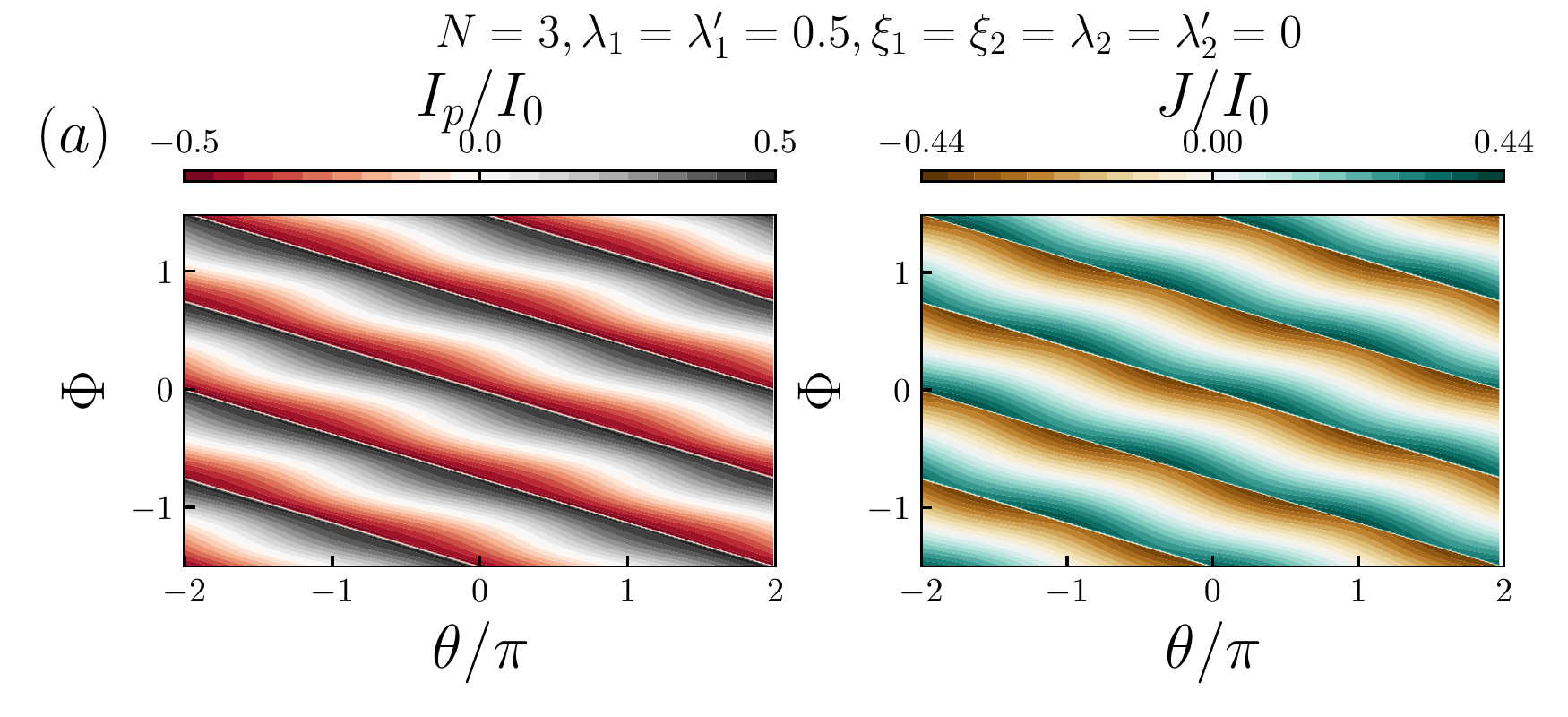}\includegraphics[scale=0.45]{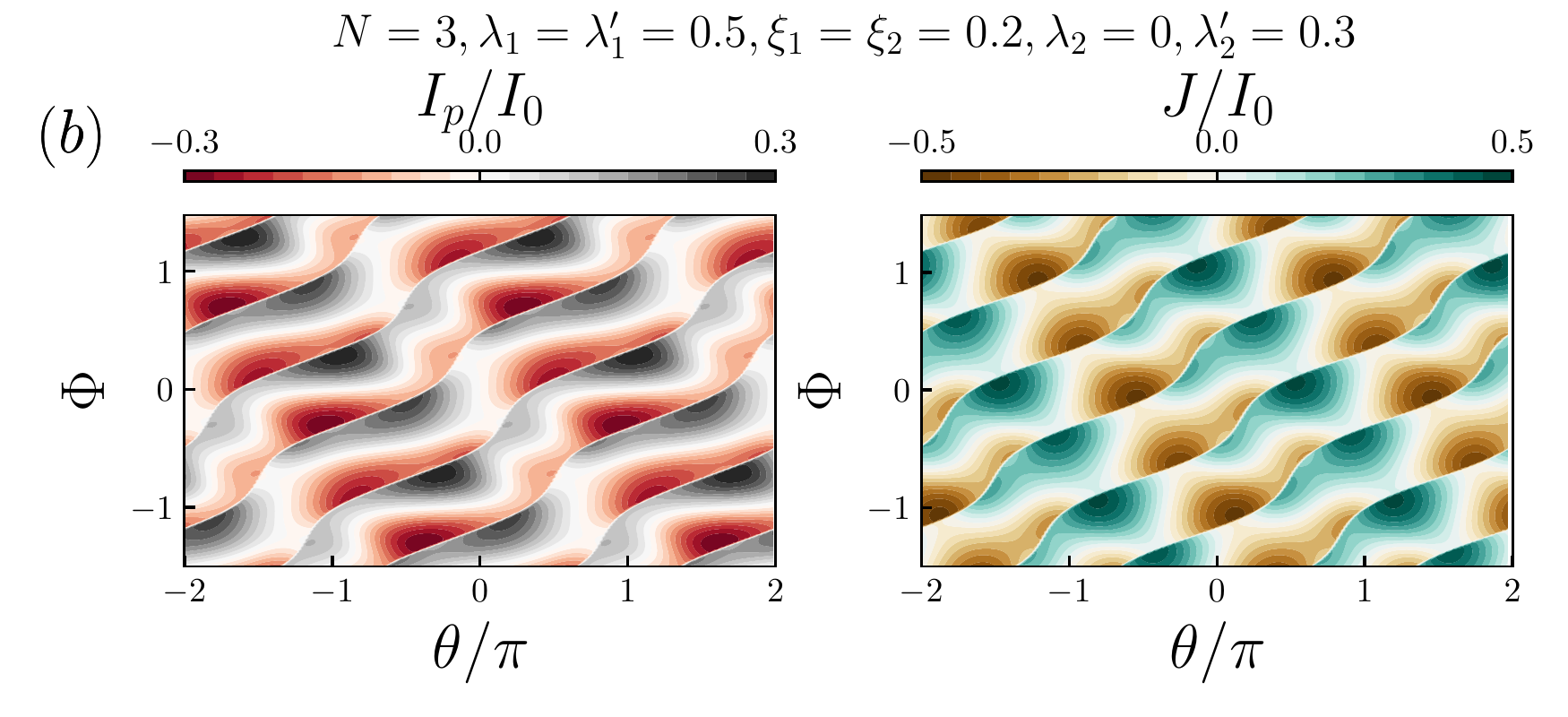}\\
	\includegraphics[scale=0.45]{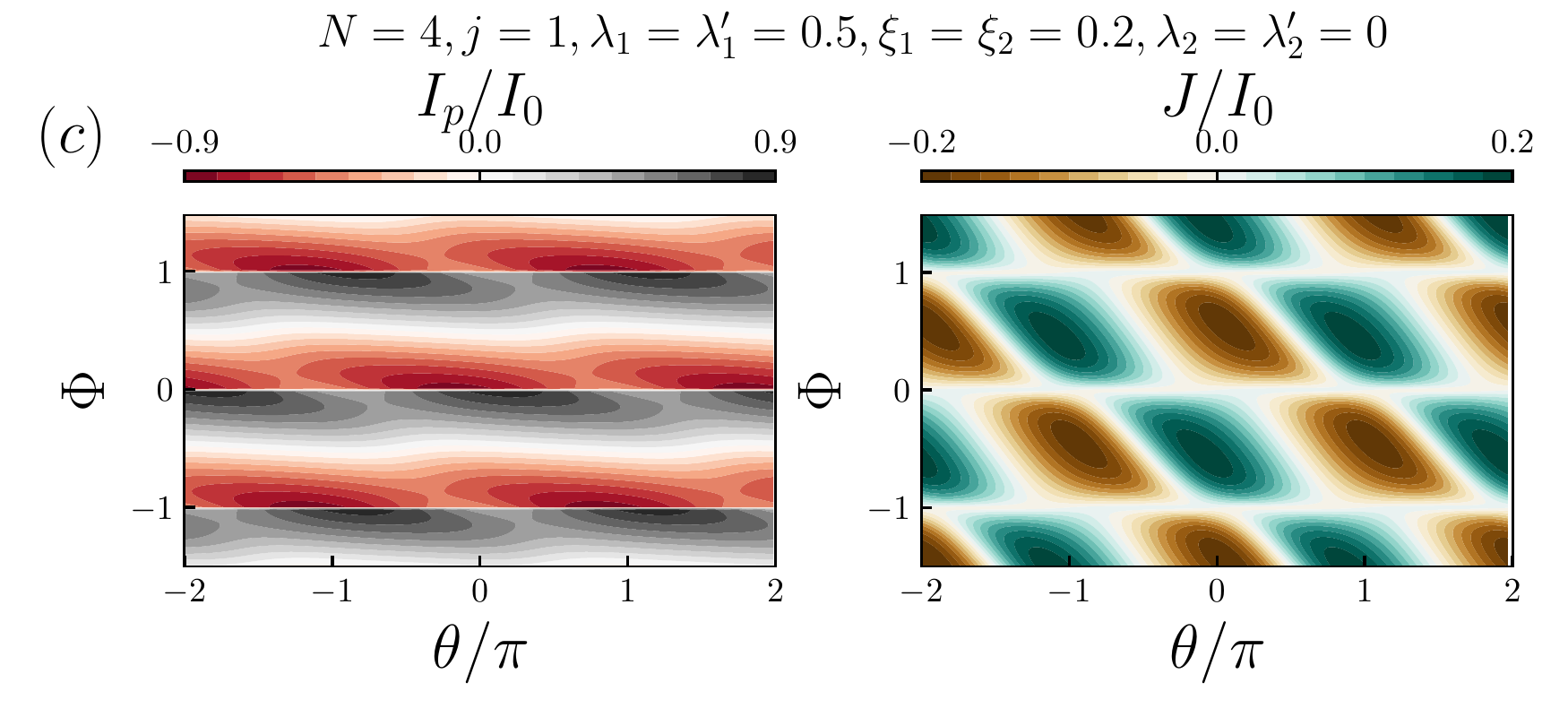}\includegraphics[scale=0.45]{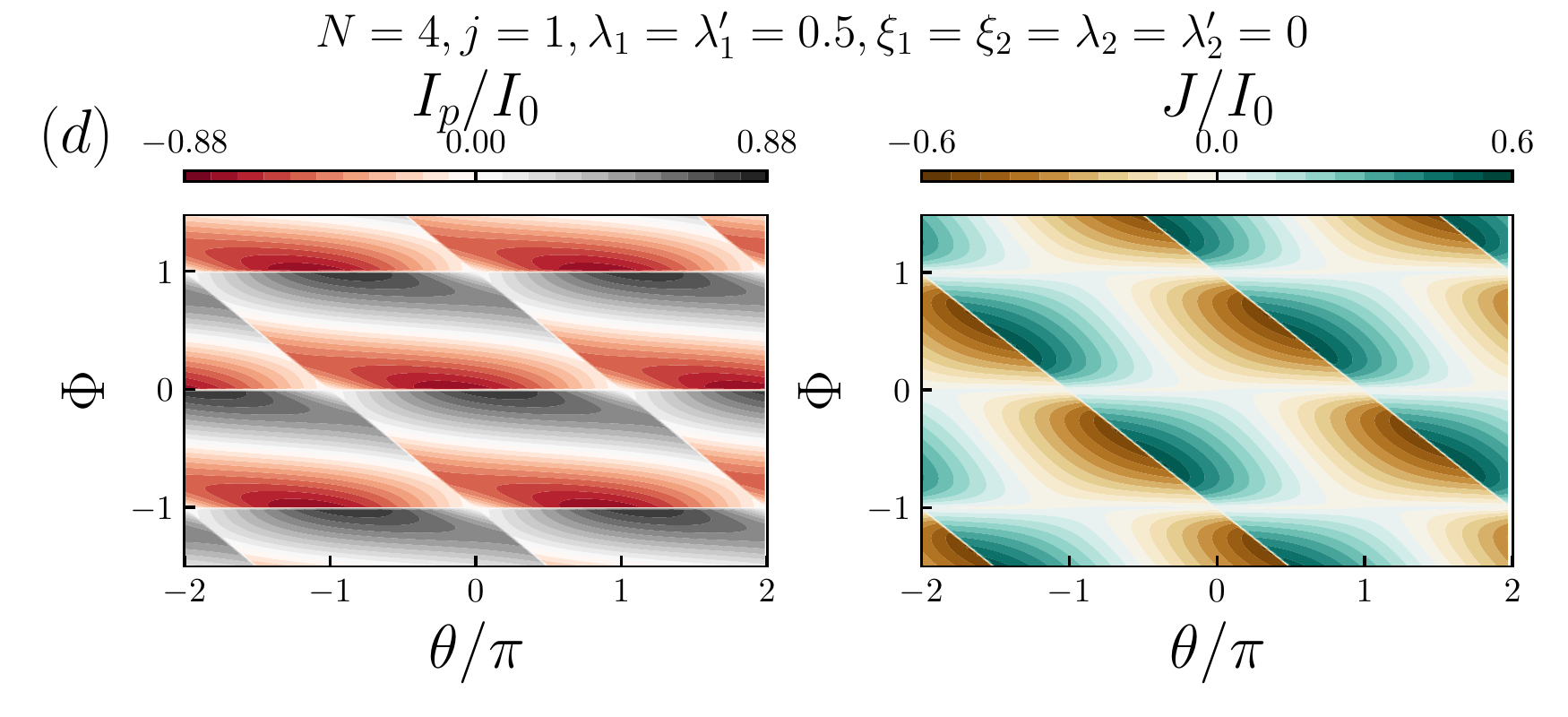}\\
	\includegraphics[scale=0.45]{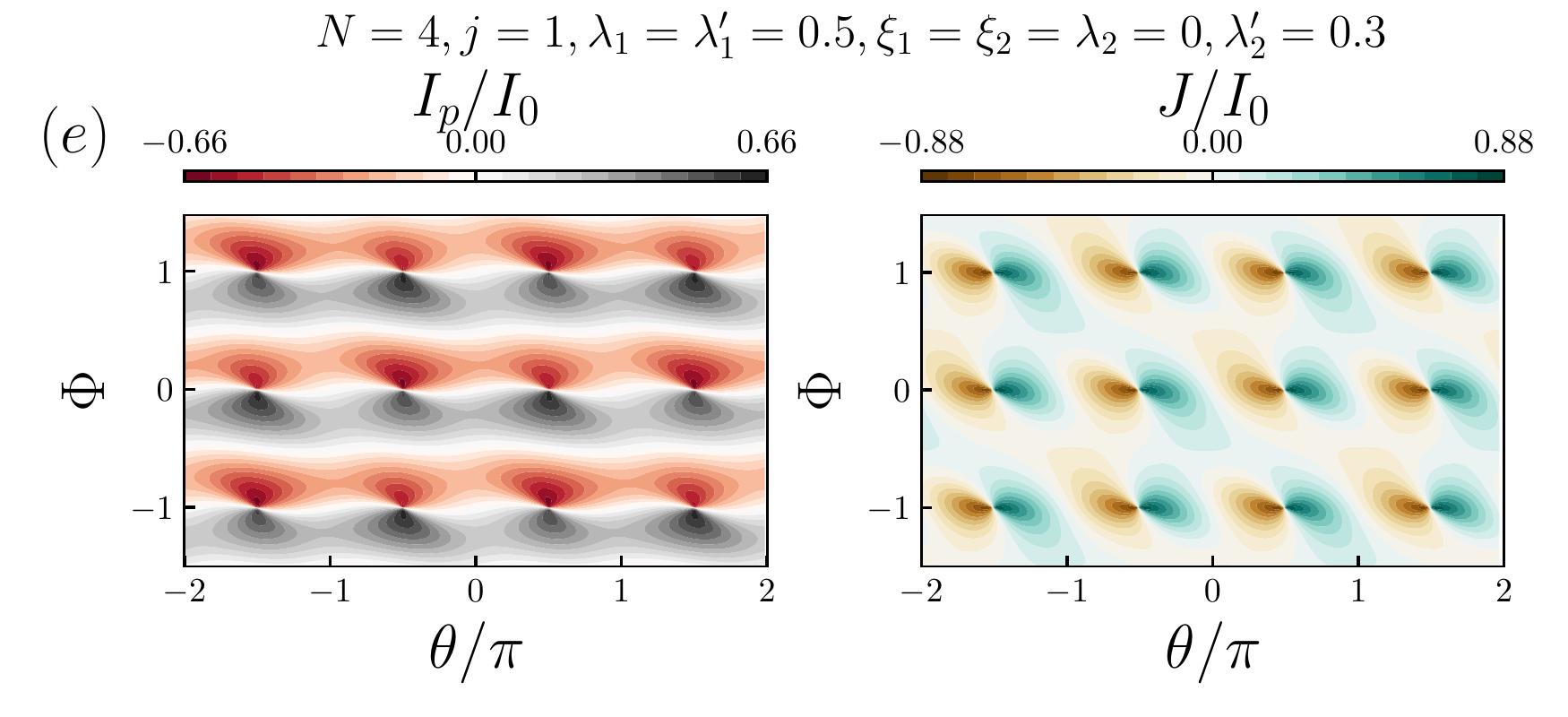}\includegraphics[scale=0.45]{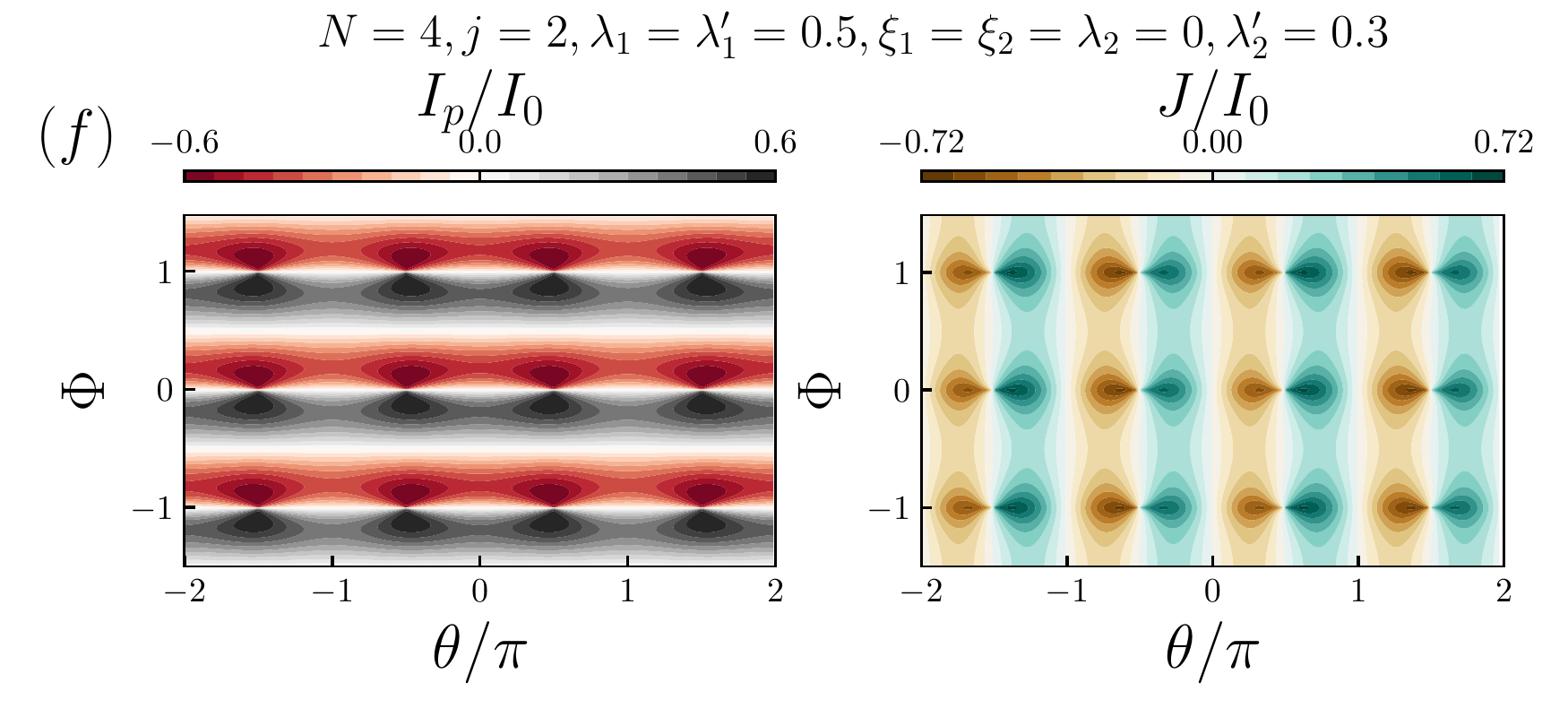}
	\caption{\label{fig:persitent_currents} Left (red and black colors) and right (yellow and green colors) columns display persistent and dc Josephson currents, respectively. $\left(a\right)$ Current signals where linear patterns are observed in both signals. $\left(b\right)$ Current signals with undulating pattern is observed in both signals. $\left(c\right)$ Current signals. Here, dc Josephson current does not detect any zero energy crossing. $\left(d\right)$ Current signals, where persistent current detects the rectangular pattern while dc Josephson just detects the pattern partially. $\left(e\right)$ Current signals, where Point pattern can be observed in both current signals. $\left(f\right)$ Current signals, where point pattern is also observed in both signals.}
\end{figure*}

The topological patterns described in the previous section, for each junction, are expected to be observed or measured experimentally through the discontinuities of persistent and dc Josephson currents. Therefore, in this section, we evaluate these two current signals for each of the topological patterns displayed in the parameter space. Hence, the persistent current in the quantum-ring at zero temperature is obtained, given by:
\begin{equation}
I_{p}/I_{0} = -\sum_{n}\frac{\partial}{\partial \Phi}E_{n}\left(\Phi,\theta\right)\,,
\label{eq:persistent_current}
\end{equation}
where $E_{n}\left(\Phi,\theta\right)$ is a dimensionless quantity (measured in units of $2t$), with particle-hole constraint in the system. Hence, we are taking the positive energy states as the occupied states, counting from zero energy. Furthermore, the persistent current is taken in units of  $I_{0} = 2t e/\hbar$.

Now if we take the differentiation of the energies with respect to the superconducting phase difference $\theta$,  we obtain the dc Josephson current at zero temperature, given by:
\begin{equation}
J/I_{0} =  2\pi\sum_{n}\frac{\partial}{\partial \theta} E_{n}\left(\Phi,\theta\right)\,.
\label{eq:Josephson_current}
\end{equation}

\begin{figure*}[tbph]
	\centering
	\includegraphics[scale=0.5]{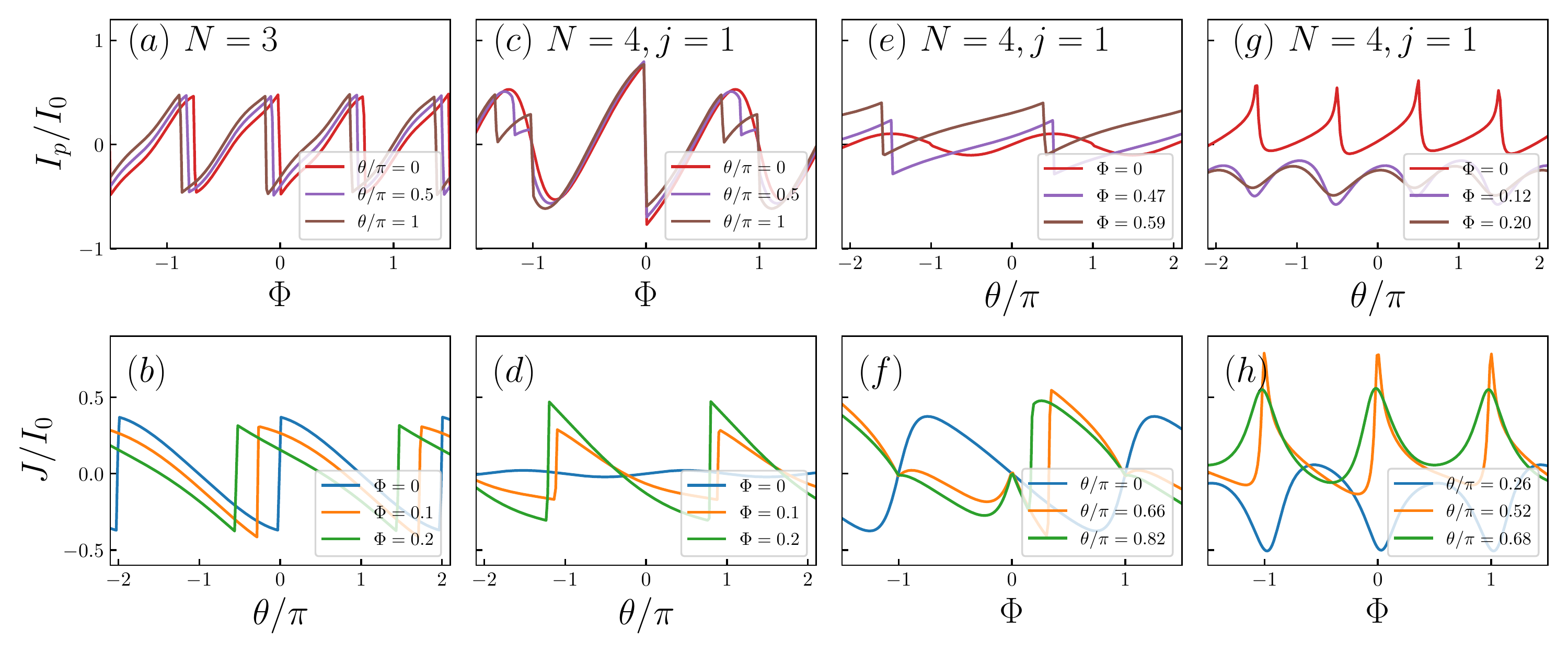}
	\caption{\label{fig:2DCurrents}  $\left(a\right)$ and $\left(c\right)$ Persistent currents as functions of magnetic flux. $\left(b\right)$ and $\left(d\right)$ dc Josephson currents, where all panels above are for $\lambda_{1} = \lambda'_{1} = 0.5, \lambda_{2} = \lambda'_{2} = 0, \xi_{1} = 0, \xi_{2} = 0$. $\left(e\right)$ and $\left(f\right)$ are persistent and dc Josephson currents as functions of the phase difference and magnetic flux, respectively. $\left(g\right)$ and $\left(h\right)$ are persistent and dc Josephson currents as functions of the phase difference and magnetic flux. Colors in labels represent fixed magnetic fluxes and phase differences.}
\end{figure*}

Here, fermion parity (FP) switching plays an important role for both types of current signals. In the spectra, we have two even and odd FP sectors that belong to positive and negative energies, or vise versa. When a pair of levels crosses zero energy, the energy excitation of the Bogoliubov quasiparticle changes sign, i.e. the ground state of the quantum-ring changes from even/odd to odd/even FP. Currents are continuous signals except for high symmetry points. Hence, when negative and positive energy modes cross at zero energy, a discontinuity in the persistent current is observed. Figure \ref{fig:persitent_currents} shows the contour plots for persistent (red and gray colors) and Josephson currents (brown and blue-green colors) related to linear, undulated, four-fold linear, rectangular, point and finally point and line node patterns in panels $(a)$, $(b)$, $(c)$, $(d)$, $(e)$ and $(f)$ respectively. Note that in $(a)$ and $(b)$ line nodes are shown in the persistent and dc Josephson currents by the sudden changes in the current signal. In $(c)$ the persistent currents show clearly the line nodes that are located at integer values of magnetic flux. On the other hand, dc Josephson current cannot detect the line nodes in this configuration since the formation of these nodes is independent of the phase difference, contrary to what happens in the previous case. In $(d)$, persistent current shows the square pattern, but dc Josephson currents only detect part of the rectangular pattern. In $(e)$, the signatures for point nodes are dipole formations in the parameter space. For persistent current, the dipole orientation is in $\Phi$, and for dc Josephson current, the orientation is in $\theta$. In panel $(f)$ point nodes can be distinguished in the same way, but since line nodes are independent of $\theta$, they cannot be distinguished by dc Josephson signals, only by persistent currents. 

A characteristic that deserves some attention can be observed in panels $(a)$ and $(b)$ from Figure \ref{fig:persitent_currents}, where persistent and dc Josephson current do not cancel out at zero magnetic flux and zero phase difference, respectively. From the spectra described above and expressions (\ref{eq:persistent_current}) and (\ref{eq:Josephson_current}), we can see precisely these phenomena. In the former, the presence of a superconducting phase difference breaks $\mathcal{T}$ symmetry generating a shift in the current signal observing persistent current at zero magnetic flux, and in the latter while the system is at equilibrium, meaning superconducting phase difference is zero, magnetic flux induces dc Josephson current. This phenomenon is not unique due to the presence of MBSs since persistent currents without gauge field can be obtained by reservoir engineering, as has been studied in \cite{keck2018persistent}. Also, in \cite{li2007persistent} a triple-arm Aharonov-Bohm ring is studied with an impurity where the breaking of $\mathcal{T}$ symmetry not only occurs by means of magnetic flux, but also due to charge stored in the three arms.

 Figure \ref{fig:2DCurrents} panel $\left(a\right)$, shows persistent currents as red, purple and brown lines that are the currents for different values of phase difference, and panel $\left(b\right)$, shows dc Josephson currents where blue, orange, and green lines are currents for different values of magnetic flux. Similar color configurations are used for the other panels. The phenomena that we mentioned above can be appreciated in the last four panels $\left(e\right)$ - $\left(h\right)$, where persistent currents are presented as a function of superconducting phase difference and the dc Josephson currents are plotted as a function of the magnetic flux. Hence, it is worth to mention that it is possible to maneuver the persistent currents with the superconducting phase differences and the dc Josephson current with the magnetic flux. In other words, our results show that in our system the persistent currents can be obtained for zero magnetic flux and the dc Josephson currents can be obtained for zero phase difference between both TSCs.     

At this point, it is important to highlight that we have checked our results by using an alternative form to calculate both currents, which are obtained by means of the variation of the total number of particles in the quantum ring, given by:

\begin{equation}
\frac{dN}{dt} = -\frac{i}{\hbar}\left[c^{\dagger}_{l'}c_{l'},H\right]\,,
\end{equation}

with the current being defined as $I = e\langle\dot{N}\rangle$. Therefore, we can express it as $I = I_{p} + J$. Hence, persistent current in the position representation reads

\begin{equation}
I_{p} =-i\frac{e}{\hbar} \sum_{l}\left(e^{i\frac{2\pi}{N}}\langle c^{\dagger}_{m}c_{m+1}\rangle - e^{-i\frac{2\pi}{N}\Phi}\langle c^{\dagger}_{m+1}c_{m}\rangle\right)\,.
\end{equation}

Through a Fourier transformation and the implementation of the fluctuation-dissipation theorem, we can write the total current as  

\begin{equation}
I = \int d\omega f\left(\omega\right)\left[j_{p}\left(\omega\right) + j_{J}\left(\omega\right)\right]\,,
\end{equation}
with $j_{p}\left(\omega\right)$ being the current density for the persistent current contribution given by \cite{maiti2011distribution}

\begin{equation}
j_{p}\left(\omega\right) = \frac{e}{\hbar\pi}\sum_{k}2\sin\left(\frac{2\pi}{N}\Phi + k\right)\text{Im}\langle\langle c^{\dagger}_{k};c_{k}\rangle\rangle^{r}_{\omega}\,,
\end{equation}
which is completely equivalent to expression (\ref{eq:persistent_current}). Hence, $j_{J}\left(\omega\right)$ is the Josephson density current, given by

\begin{align}
j\left(\omega\right) = -\frac{4e}{h}\sum_{k}\text{Im}[&\lambda'^{*}_{\left(-\right)}e^{ikj}\text{Im}\langle\langle a^{\dagger};c^{\dagger}_{k} \rangle\rangle^{r}_{\omega} +\\ &\lambda'_{\left(+\right)}e^{ikj}\text{Im}\langle\langle a; c^{\dagger}_{k}\rangle\rangle^{r}_{\omega} +\nonumber \\ & \lambda_{\left(-\right)}\text{Im}\langle\langle f^{\dagger};c^{\dagger}_{k}\rangle\rangle^{r}_{\omega} + \lambda_{\left(+\right)}\text{Im}\langle\langle f;c^{\dagger}_{k}\rangle\rangle^{r}_{\omega} ]\,, \nonumber 
\end{align}
where,

\begin{align*}
 \lambda'_{\left(-\right)} =& \frac{1}{2} \left(\lambda'_{1}e^{-i\frac{\theta}{2}} - \lambda'_{2}e^{i\frac{\theta}{2}}\right),\\
 \lambda'_{\left(+\right)} =& \frac{1}{2}\left(\lambda'_{1}e^{i\frac{\theta}{2}} + \lambda'_{2}e^{-i\frac{\theta}{2}}\right),\\
 \lambda_{\left(\pm\right)} =& \frac{1}{2}\left(\lambda_{1} \pm \lambda_{2}\right).
\end{align*}

\section{Summary} \label{Sec. 5}

In this work, we studied a system formed by a quantum-ring embedded between two TSCs, both hosting MBSs at their ends. A magnetic flux crosses the ring, and we considered two quantum-ring sizes, i.e., $N=3$ and $N=4$. Based on the tuning of the magnetic flux and the superconducting phase difference, we obtain the spectra of the system, observing zero energy-crossing points and flat bands. All the features obtained in the spectra could be observed through persistent and dc Josephson current measurements, by looking for the discontinuities in the signals for linear and undulated patterns. For the case of point nodes, the characteristic is the formation of the dipole-like structure in the currents.  Interestingly, our results show that the persistent current and the dc Josephson current can be manipulated by variations in superconducting phase and magnetic flux, respectively, being clear that these phenomena are not due to the presence of MBSs. Thus, we believe our findings can be helpful to understand the interplay between both current signals and their relation with the system spectra.   

Furthermore, we highlight that the currents obtained may be accessible in experiments. The most common setups for the detection of MBSs in 1D TSCs with QD are made of InAs and GaAs \cite{deng2018nonlocality,deng2016majorana,prada2017measuring}. In these setups, a segment of nanowire is covered with Al to build a TSC, leaving bare a shorter section of a few $nm$ at the edges. These bare sections play the role of QDs where the charge is controlled by gate voltages. Based on the above description, a similar engineering can be used to make a quantum-ring with two TSCs. The QDs can be inserted as local impurities in the ring.
Given the fact that nanowires usually have a width in the order of $nm$ and lengths in the order of $\mu m$ \cite{vaitiekenas2020flux}, the actual ring diameter could be also in the order of $\mu m$, which can hold persistent currents with amplitudes $\sim 1 nA$ at temperatures below $1K$. A wide range of magnetic fields can be experimentally obtained by the implementation of a single-crystal Si cantilever \cite{bleszynski2009persistent}.
As a starting point, the experimental realization for the measurement of dc Josephson currents in multiterminal TSCs could be based on the promising experiments in multiterminal Josephson junctions with conventional superconductivity, where the superconductors are made of Al on top of an InAs 2DEG \cite{pankratova2020multiterminal,draelos2019supercurrent,graziano2020transport}. 

Finally, the configuration gives the possibility of using our proposal in braiding operations if we consider a time-dependent magnetic flux. In \cite{PhysRevLett.126.117701}, six possible braiding operations are shown for four Majorana modes. In our proposal, a resonant manipulation can be performed by applying linear time-dependent magnetic flux pulses, in order to produce oscillating coupling with the frequency, matching the energy spacing between the Majorana modes.

\begin{table*}[ht]
	\begin{center}
		\begin{tabular}{|c|c|c|c|c|c|c|c|}
			\hline
			Junction & $\lambda_{1}$ & $\lambda_{2}$ & $\lambda_{1}'$ & $\lambda_{2}'$ & $\xi_{1}$ & $\xi_{2}$& Nodes\\
			
			\hline
			
			\multirow{3}{*}{$A$} & \multirow{3}{*}{$\times$} &  & \multirow{3}{*}{$\times$} &  & 0 & 0 &  Line\\
			&  &  &  &  & $\times$ & 0 & Flat band\\
			& & & & & $\times$ & $\times$ & Line\\
			
			\hline
			
			\multirow{1}{*}{$B$} & $\times$ & $\times$ & $\times$ &  & $\times$ & 0 & Flat band \\
			
			\hline
			
			\multirow{3}{*}{$C$} & \multirow{3}{*}{$\times$} &  & \multirow{3}{*}{$\times$} & \multirow{3}{*}{$\times$} & $\times$ & 0 & Line\\
			&  &  &  &  & 0 & $\times$ & Flat band\\
			& & & & & $\times$ & $\times$ & Undulate\\
			
			\hline
			
			\multirow{4}{*}{$D$} & \multirow{4}{*}{$\times$} & \multirow{4}{*}{$\times$} & \multirow{4}{*}{$\times$} & \multirow{4}{*}{$\times$} & 0 & 0 & Line\\
			&  & &  &  & $\times$ & 0 & Undulate\\
			& & & & & 0 & $\times$ & Line\\
			& & & & & $\times$ & $\times$ & Undulate\\ 
			\hline
		\end{tabular}
		\caption{\label{table1} List of topological patterns for $N = 3$ quantum-ring in a two-junction with 1D TSCs. The first column indicates the type of junction. The next columns represent the non-zero couplings marked with an $\times$. The latter column shows the topological pattern observed in the spectra or the presence of flat bands.}
	\end{center}
\end{table*}

\begin{table*}[ht]
	\begin{center}
		\begin{tabular}{|c|c|c|c|c|c|c|c|c|}
			\hline
			Junction & j & $\lambda_{1}$ & $\lambda_{2}$ & $\lambda_{1}'$ & $\lambda_{2}'$ & $\xi_{1}$ & $\xi_{2}$ & Nodes\\
			\hline
			\multirow{6}{*}{$A$} & \multirow{4}{*}{1} & \multirow{6}{*}{$\times$} &  & \multirow{6}{*}{$\times$} & & 0 & 0 & Rectangular\\
			& & & & & & 0 & $\times$ & Line and Flat band\\
			& & & & & & $\times$ & 0 & Line and Flat band\\
			& & & & & & $\times$ & $\times$ & Four fold line\\
			\cline{2-2}
			\cline{7-9}
			& \multirow{3}{*}{2} & & & & & 0 & $\times$ & Line\\
			& & & & & & $\times$ & 0 & Line\\
			& & & & & & $\times$ & $\times$ & Four fold line\\
			\hline
			\multirow{4}{*}{$B$} & \multirow{2}{*}{1} & \multirow{4}{*}{$\times$} & \multirow{4}{*}{$\times$} & \multirow{4}{*}{$\times$} & & $\times$ & 0 & Flat band\\
			& & & & & & $\times$ & $\times$ & Line\\
			\cline{2-2}
			\cline{7-9}
			&\multirow{2}{*}{2} & & & & & $\times$ & 0 & Line\\
			& & & & & &$\times$ & $\times$ & Line\\
			\hline
			\multirow{7}{*}{C} & \multirow{4}{*}{1} & \multirow{7}{*}{$\times$} & & \multirow{7}{*}{$\times$} & \multirow{7}{*}{$\times$} & 0 & 0 & Point and Flat band\\
			& & & & & & 0 & $\times$ & Point and Flat band\\
			& & & & & & $\times$ & 0 & Point and Flat band\\
			& & & & & & $\times$ & $\times$ & Point and Line\\
			\cline{2-2}
			\cline{7-9}
			& \multirow{3}{*}{2} & & & & & $\times$ & 0 & Point\\
			& & & & & & 0 & $\times$ & Line\\
			& & & & & & $\times$ & $\times$ & Point and Line\\
			\hline
			\multirow{5}{*}{$D$} & \multirow{3}{*}{1} & \multirow{5}{*}{$\times$} & \multirow{5}{*}{$\times$} & \multirow{5}{*}{$\times$} & \multirow{5}{*}{$\times$} & $\times$ & 0 & Line \\
			& & & & & & 0 & $\times$ & Line \\
			& & & & & & $\times$ & $\times$ & Line\\
			\cline{2-2}
			\cline{7-9}
			& \multirow{2}{*}{2} & & & & & $\times$ & 0 & Point\\
			& & & & & & $\times$ & $\times$ & Point and Line\\
			\hline
		\end{tabular}
		\caption{\label{table2} Topological patterns for $N = 4$ quantum-ring in a two-junction with 1D TSCs. The columns represent the same as Table 1 with a distinction in the second column from this table which represents the position of one TSC at sites $j = 1,2$ while the other remains fixed at $j = 0$.}
	\end{center}
\end{table*}

\appendix
\section*{Appendix A: Hamiltonian Symmetry.}

The Hamiltonian of the system in Majorana representation, and in a BdG form, takes the form of a skew-symmetric matrix. Considering a single mode in the quantum ring, the spinor is written as $\Psi = \left(\alpha^{\left(1\right)}_{0}\alpha^{\left(2\right)}_{0}\eta_{1}\eta_{2}\eta_{3}\eta_{4}\right)^{T}$ and the Hamiltonian can be written as:
\begin{widetext}
\begin{equation*}
    H^{BdG}_{R} = \frac{1}{2}\epsilon\left(\Phi\right)
    \begin{pmatrix}
    0 & i & 0 & 0 & 0 & 0\\ 
    -i & 0 & 0 & 0 & 0 & 0\\
    0 & 0 & 0 & 0 & 0 & 0\\
    0 & 0 & 0 & 0 & 0 & 0\\
    0 & 0 & 0 & 0 & 0 & 0\\
    0 & 0 & 0 & 0 & 0 & 0
    \end{pmatrix},
\end{equation*}
\begin{equation*}
    H^{BdG}_{TSC} = \frac{1}{2}\xi_{1}
    \begin{pmatrix}
    0 & 0 & 0 & 0 & 0 & 0\\
    0 & 0 & 0 & 0 & 0 & 0\\
    0 & 0 & 0 & i & 0 & 0\\
    0 & 0 & -i & 0 & 0 & 0\\
    0 & 0 & 0 & 0 & 0 & 0\\
    0 & 0 & 0 & 0 & 0 & 0
    \end{pmatrix} + \frac{1}{2}\xi_{2}
    \begin{pmatrix}
    0 & 0 & 0 & 0 & 0 & 0\\
    0 & 0 & 0 & 0 & 0 & 0\\
    0 & 0 & 0 & 0 & 0 & 0\\
    0 & 0 & 0 & 0 & 0 & 0\\
    0 & 0 & 0 & 0 & 0 & i\\
    0 & 0 & 0 & 0 & -i & 0
    \end{pmatrix},
\end{equation*}
\begin{align*}
    H^{BdG}_{C} = \frac{1}{2}\lambda_{1}
    \begin{pmatrix}
    0 & 0 & 0 & 0 & 0 & 0\\
    0 & 0 & -i & 0 & 0 & 0\\
    0 & i & 0 & 0 & 0 & 0 \\
    0 & 0 & 0 & 0 & 0 & 0 \\
    0 & 0 & 0 & 0 & 0 & 0 \\
    0 & 0 & 0 & 0 & 0 & 0
    \end{pmatrix}
    +& \frac{1}{2}\lambda_{2}
    \begin{pmatrix}
    0 & 0 & 0 & i & 0 & 0\\
    0 & 0 & 0 & 0 & 0 & 0\\
    0 & 0 & 0 & 0 & 0 & 0 \\
    -i & 0 & 0 & 0 & 0 & 0 \\
    0 & 0 & 0 & 0 & 0 & 0 \\
    0 & 0 & 0 & 0 & 0 & 0
    \end{pmatrix}\\
    +\frac{1}{2}\lambda'_{1}S\left(\theta\right)
    \begin{pmatrix}
    0 & 0 & 0 & 0 & i & 0\\
    0 & 0 & 0 & 0 & 0 & 0\\
    0 & 0 & 0 & 0 & 0 & 0 \\
    0 & 0 & 0 & 0 & 0 & 0 \\
    -i & 0 & 0 & 0 & 0 & 0 \\
    0 & 0 & 0 & 0 & 0 & 0
    \end{pmatrix}
    +& \frac{1}{2}\lambda'_{1}C\left(\theta\right)
    \begin{pmatrix}
    0 & 0 & 0 & 0 & 0 & 0\\
    0 & 0 & 0 & 0 & -i & 0\\
    0 & 0 & 0 & 0 & 0 & 0 \\
    0 & 0 & 0 & 0 & 0 & 0 \\
    0 & i & 0 & 0 & 0 & 0 \\
    0 & 0 & 0 & 0 & 0 & 0
    \end{pmatrix}\\
    +\frac{1}{2}\lambda'_{2}C\left(\theta\right)
    \begin{pmatrix}
    0 & 0 & 0 & 0 & 0 & i\\
    0 & 0 & 0 & 0 & 0 & 0\\
    0 & 0 & 0 & 0 & 0 & 0 \\
    0 & 0 & 0 & 0 & 0 & 0 \\
    0 & 0 & 0 & 0 & 0 & 0 \\
    -i & 0 & 0 & 0 & 0 & 0
    \end{pmatrix}
    +& \frac{1}{2}\lambda'_{2}S\left(\theta\right)
    \begin{pmatrix}
    0 & 0 & 0 & 0 & 0 & 0\\
    0 & 0 & 0 & 0 & 0 & -i\\
    0 & 0 & 0 & 0 & 0 & 0 \\
    0 & 0 & 0 & 0 & 0 & 0 \\
    i & 0 & 0 & 0 & 0 & 0 \\
    0 & 0 & 0 & 0 & 0 & 0
    \end{pmatrix}.
\end{align*}
\end{widetext}
It is straightforward to distinguish the basis on which the Hamiltonian is written. On this basis, and for a single mode in the ring, the Hamiltonian is expressed in terms of some elements of the $SU\left(6\right)$ Lie algebra in $\lambda$-representation. Hence, we can write the Hamiltonian as
\begin{align}
    \begin{split}
    H_{BdG} &= \frac{1}{2}\left(-\epsilon\left(\Phi\right)\tilde{\lambda}_{2} - \xi_{1}\tilde{\lambda}_{14} - \xi_{2}\tilde{\lambda}_{34} + \lambda_{1}\tilde{\lambda}_{17}\right.\\
    & + \lambda_{1}\tilde{\lambda}_{7} - \lambda_{2}\tilde{\lambda}_{10} -\lambda'_{1}S\left(\theta\right)\tilde{\lambda}_{17} + \lambda'_{1}C\left(\theta\right)\tilde{\lambda}_{19}\\
    & \left.- \lambda'_{2}C\left(\theta\right)\tilde{\lambda}_{26} + \lambda'_{2}S\left(\theta\right)\tilde{\lambda}_{28}\right),
    \end{split}
\end{align}
where $\epsilon\left(\Phi\right) = \cos\left(2\pi\Phi/N\right)$, $S\left(\theta\right) = \sin\left(\theta/2\right)$, $C\left(\theta\right) = \cos\left(\theta/2\right)$ and $\tilde{\lambda}_{i}$ are the imaginary elements of the $SU\left(6\right)$ group. Therefore, we can define a vector $\mathbf{h}$ with a dimension thirty-six with non-zero components written in the above Hamiltonian. Similarly, we can represent as a vector the whole basis of the $SU\left(6\right)$ algebra and the full Hamiltonian, written in a compact form as $H_{BdG} = \mathbf{h}\cdot\mathbf{\Lambda}/2$. This is generalized easily for any number of modes in the quantum-ring. Since the Hamiltonian can be written as a linear combination of the basis elements of the $SU\left(2N + 4\right)$, the system possesses that symmetry. 

\section*{Acknowledgements}
F.G.M. is grateful for the funding of CONICYT-Chile scholarship No. 21170550. J.P.R.-A is grateful for the funding of FONDECYT Postdoc. Grant No. 3190301 (2019). L.R. and P.A.O. acknowledges support from FONDECYT Grant No. 1180914 and 1201876.

\bibliographystyle{apsrev4-1}
\bibliography{biblio}

\end{document}